\documentclass[10pt,journal,a4paper]{IEEEtran}
\usepackage{amsmath}
\usepackage{amsthm}
\usepackage{amsfonts}
\usepackage{amssymb}
\usepackage{makeidx}
\usepackage{cite}
\usepackage{graphicx,bigints}
\usepackage{mathtools}
\usepackage{cases}
\usepackage{color}
\usepackage{hyperref}
\usepackage{bbm}
\usepackage[normalem]{ulem}
\usepackage{braket}
\usepackage{textcomp,gensymb}
\usepackage{tablefootnote}
\usepackage{soul}

\usepackage{enumitem}

\usepackage{epstopdf}
\usepackage{xcolor}
\usepackage{lipsum}

\usepackage{multirow}

\usepackage[makeroom]{cancel}

\makeatletter
\DeclareFontFamily{U}{tipa}{}
\DeclareFontShape{U}{tipa}{m}{n}{<->tipa10}{}
\newcommand{\arc@char}{{\usefont{U}{tipa}{m}{n}\symbol{62}}}%

\newcommand{\arc}[1]{\mathpalette\arc@arc{#1}}

\newcommand{\arc@arc}[2]{%
	\sbox0{$\m@th#1#2$}%
	\vbox{
		\hbox{\resizebox{\wd0}{\height}{\arc@char}}
		\nointerlineskip
		\box0
	}%
}
\makeatother

\newtheorem{theorem}{Theorem}

\newtheorem{lemma}{Lemma}

\newtheorem{example}{Example}

\newtheorem{remark}{Remark}

\DeclareMathOperator{\cO}{\mathcal{O}}

\DeclareMathOperator{\cR}{\mathcal{R}}

\DeclareMathOperator{\SNR}{\textrm{SNR}}

\DeclareMathOperator{\bR}{\mathbb{R}}

\DeclareMathOperator{\bN}{\mathbb{N}}
\DeclareMathOperator{\bP}{\mathbf{P}}
\DeclareMathOperator{\ind}{\mathbbm{1}}
\DeclareMathOperator{\bE}{\mathbf{E}}
\DeclareMathOperator{\bZ}{\mathbb{Z}}
\newcommand*\diff{\mathop{}\!\mathrm{d}}

\newcommand*\nnb{\nonumber}

\newcommand\independent{\protect\mathpalette{\protect\independenT}{\perp}}
\def\independenT#1#2{\mathrel{\rlap{$#1#2$}\mkern2mu{#1#2}}}

\definecolor{sandy}{HTML}{E6E2AF}
\definecolor{stone}{HTML}{A7A37E}
\definecolor{beach}{HTML}{EFECCA}
\definecolor{ocean}{HTML}{046380}
\definecolor{diver}{HTML}{002F2F}

\definecolor{Firenze1}{HTML}{468966}
\definecolor{Firenze2}{HTML}{FFF0A5}
\definecolor{Firenze3}{HTML}{FFB03B}
\definecolor{Firenze4}{HTML}{B64926}
\definecolor{Firenze5}{HTML}{8E2800}
\definecolor{mediumpersianblue}{rgb}{0.0, 0.4, 0.65}
\definecolor{hongik}{HTML}{004498}
\definecolor{newhongik}{HTML}{0000EE}
\definecolor{cobalt}{rgb}{0.0, 0.28, 0.67}
\definecolor{burntorange}{rgb}{0.8, 0.33, 0.0}

\definecolor{ultramarineblue}{rgb}{0.25, 0.4, 0.96}

\title{Stochastic Geometry Analysis of RIS-Assisted Cellular Networks with Reflective Intelligent Surfaces on Roads}

\author{Chang-Sik Choi, Junhyeong Kim, and Junil Choi
	\IEEEcompsocitemizethanks{\IEEEcompsocthanksitem{Chang-Sik Choi is with Hongik University, South Korea (email: chang-sik.choi@hongik.ac.kr) Junil Choi is with Kaist, South Korea. Junhyeong Kim is with ETRI, South Korea. Last revised: \today}}
}
\begin{document}
	\maketitle 
	
	\begin{abstract}
Reconfigurable intelligent surfaces (RISs) provide alternative routes for reflected signals to network users, offering numerous applications. This paper explores an innovative approach of strategically deploying RISs along road areas to leverage various propagation and blockage conditions present in cellular networks with roads. To address the local network geometries shown by such networks, we use a stochastic geometry framework, specifically the Cox point processes, to model the locations of RISs and vehicle users. Then, we define the coverage probability as the chance that either a base station or an RIS is in line of sight (LOS) of the typical user and that the LOS signal has a signal-to-noise ratio (SNR) greater than a threshold. We derive the coverage probability as a function of key parameters such as RIS density and path loss exponent. We observe that the network geometry highly affects the coverage and that the proposed RIS deployment effectively leverages the underlying difference of attenuation and blockage, significantly increasing the coverage of vehicle users in the network. With experimental results addressing the impact of key variables to network performance, this work serves as a versatile tool for designing, analyzing, and optimizing RIS-assisted cellular networks with many vehicles.
	\end{abstract}


\section{Introduction}

\IEEEPARstart{R}{IS}s are considered one of the key enablers for various applications in future wireless technologies \cite{9424177}. These intelligent surfaces modify the propagation directions of wireless signals, providing alternate paths to reach various users in a wider area \cite{8796365,9360709}. In particular, research and experiments have identified that the locations and surroundings of wireless transmissions dictate the performance of the RIS-enabled communication networks \cite{9424177,8796365,9360709,8424015,9237116,38901}. For example, users in dense urban areas may not get enough benefits from RISs because of high facade buildings creating significant signal attenuation, whereas vehicle users on roads may benefit more from RISs because there are relatively small obstacles on roads compared to dense urban areas, resulting in modest attenuation \cite{8424015,9237116,38901}. Provided that there are various types of users in modern wireless networks and that the geometry of the network determines the performance of RIS-assisted networks, the performance of an RIS-enabled architecture needs to be analyzed under an accurate modeling of the network geometry that base stations, users, and RISs jointly create.

By strategically deploying RISs solely near the road infrastructure, this paper provides a novel approach to addressing and leveraging the unique network geometry of RIS-assisted cellular networks with roads and vehicle users. Using stochastic geometry, we first accurately model the spatial clustering of vehicle users and RISs, then analyze the network performance by evaluating the fundamental limit of maximum connectivity that RISs would offer. This work aims to mathematically showcase that the proposed RIS deployment strategy sharply improves the fundamental connectivity of modern urban cellular networks with roads by exploiting the differences in attenuation and blockage of such networks.

\subsection{Related Work}\label{S-related}

Research on RISs has primarily focused on the practical and theoretical aspects of these metasurfaces. Some research papers have discussed the physical design and implementation. Specifically, \cite{8990007} investigated the phase control of RISs. Propagation leakage and hardware implementation issues were studied in \cite{9789438}. Complex channel models and the channel hardening effect were analyzed in \cite{9300189,10175074}. Meanwhile, \cite{9357969} examined the passive designs of RISs, while \cite{9497709} analyzed the energy efficiency of RIS-enabled networks. \cite{9831036} studied the use of RIS for mmWave vehicular networks. \cite{9918631} explored RIS-assisted mmWave communication networks, and \cite{9722711} analyzed the localization application of wireless networks assisted by RISs. \cite{9505311} investigated vehicle-to-everything communications with RISs, and \cite{9744412} discussed the use of RISs on roads by proposing enabling protocols and architecture. Note that the above studies focused on the link-level behavior of RIS-assisted networks.

In contrast to  the above work \cite{8990007,9789438,9300189,9831036,9918631,9722711,9505311,10175074,9357969,9497709,9744412}, system-level performance analysis delivers practical design principles and high-level insights by accounting for large-scale interactions of network elements. The system-level performance analysis of RIS-enabled networks has been conducted either by large-scale simulation-based experiments or by employing tools from stochastic geometry \cite{schneider2008stochastic,baccelli2010stochastic,haenggi2012stochastic,chiu2013stochastic}. In the analysis of various networks, the stochastic geometry approach exhibits several advantages over the simulation-based method. For instance, \cite{6042301} represents the network performance as key geometric variables, highlighting the impacts of those variables and producing useful large-scale optimization ideas.

Regarding the analysis of wireless networks with RISs, \cite{9110835,9224676} analyzed the coverage probability of an RIS-assisted mmWave communication network. A study on the large-scale connectivity of RIS-assisted networks was conducted in \cite{9174910,9868205}. In \cite{9174910}, the authors placed reflecting surfaces specifically on blockages and then derived the probability that a randomly located user is connected to a transmitter with the assistance of RISs. Employing a random blockage model in \cite{6290250}, the large-scale performance behavior of a RIS-assisted network is identified in \cite{9868205} as a function of parameters such as the number of RISs. It is important to mention that both papers used the random blockage model \cite{6290250} that each link is assumed to undergo independent random blockage.

Leveraging stochastic geometry and point processes, the geometric impacts of RISs on a non-orthogonal multiple access network were assessed in \cite{9606895}. It was shown by analysis in \cite{9673721} that the typical network performance is determined by the spatial interaction of network elements especially by the relative density of network users. This study reveals that not only the locations of RISs but also the relative density of users decide the large-scale performance of RIS-assisted networks.

The majority of work on RIS-assisted networks used homogeneous Poisson point processes \cite{baccelli2010stochastic} to model the locations of base stations and users. The Poisson point process modeling often provides a simple expression for the network performance since the spatial interaction between the network elements is ignored. Nevertheless, in real-world networks, spatial dependency exists not only between the base stations and users but also between the base stations and RISs, as well as their users \cite{10064007,sun2023performance}. Acknowledging the importance of such spatial dependency in the analysis of network performance, \cite{10064007} characterized the spatial correlation of base stations and RISs as a Gaussian point process. Similarly, \cite{sun2023performance} addressed the spatial dependency of RISs to base stations as a cluster point process \cite{chiu2013stochastic}. In \cite{sun2023performance}, RISs are strategically deployed close to base stations to serve users that cannot be well served by base stations alone. The results of \cite{sun2023performance} indicates that the locations of RISs must be determined strategically for the best coverage. 

In this paper, we focus on a forthcoming scenario of 5G or 6G where many vehicle users actively participate in communications. In particular, as demonstrated by \cite{9206044,38901}, we leverage the propagation characteristics of signals on roads, which differ from that of signals on urban or rural areas. In this context, we propose and analyze a strategic deployment of RISs close to road infrastructure so that RISs can leverage the favorable propagation and attenuation characteristics on roads areas. Our approach takes advantage of the local propagation and blockage characteristics of links on roads while accurately representing the spatial dependency between base stations, RISs, and users in the network using stochastic geometry. Since our work accounts for the spatial correlation of RISs and users and their distributions, it contrasts with the models in \cite{8910627,10064007}, where the spatial correlation or RISs and users were ignored for simpler analysis.

It is worthwhile to mention that some recent studies have emphasized the role of geometric clustering and spatial correlation in network performance. By focusing on the fact that vehicle users are on roads, \cite{morlot2012population,8340239,8357962,8419219,8353411,8796442,9201475,9354063,9477118,9524528,9782581,9792580,10066317} used Cox point processes to model network elements close to road infrastructure, such as vehicle users or roadside units. The use of the Cox model revealed that the performance of spatially-correlated networks is very different from the performance of networks without spatial correlation. However, no study has analyzed the performance of RIS-assisted cellular networks where RISs are deployed close to vehicle users, resulting in strong spatial correlation between RISs and vehicle users in real-world. In particular, no study has established an analytical framework to assess the effect of RISs and the links they make with vehicle users nearby and with handset users at random locations. By accounting for the spatial correlation of vehicle users and RISs, this paper builds a systematic framework based on stochastic geometry to model and analyze the fundamental connectivity of an RIS-assisted cellular network with road infrastructure.

\subsection{Theoretical Contributions}

\subsubsection{Exploiting Network Geometry with Roadside RISs}
In the context of stochastic geometry modeling and analysis, this paper explores a novel approach of strategically placing RISs on roads to significantly improve the connectivity of vehicular users. The network performance achieved by our approach is analytically evaluated under the stochastic geometry framework developed in this paper. Specifically, we model base stations as a Poisson point process. Then, we consider two types of users in the network. Handset users are modeled as an independent Poisson point process, whereas vehicle users are modeled as a stationary Cox point process that simultaneously creates lines for roads and points on the lines for vehicles. We use the Cox point process to account for the geometric fact that vehicle users are exclusively on roads, exhibiting strong correlation between roads and vehicles in real-world. Then, conditionally on the same line process, we model the locations of RISs as another independent Cox point process. Consequently, the RISs and vehicle users are spatially correlated as they are represented as points on the same lines. The developed framework not only accurately characterizes the unique spatial relationship between RISs and vehicle users but also provides a foundation for evaluating the statistics of various links between base stations, RISs, vehicle users, and handset users. Using this spatial model, the blockage profile and attenuation characteristics of these various local links are assessed in a single unified framework, producing the network performance.

Note that some studies \cite{9110835, 9224676,9174910,9868205,9606895,9673721,10064007,sun2023performance} analyzed the impacts of RISs employing stochastic geometry. Our work differs from these studies because we specifically impose spatial dependence between users and RISs by representing them as Cox point processes conditional on the same line process. Both vehicle users and RISs are exclusively on the same road infrastructure. Although the Cox model for spatial correlation of vehicles was investigated in \cite{morlot2012population,8340239,8357962,8419219,8353411,9201475,9477118,9524528,9782581,9792580}, the spatial correlation of RISs and vehicles has never been emphasized and analyzed in a single framework. We will shortly see that once RISs are deployed in proximity to vehicle users, the RISs take advantage of their local propagation and blockage characteristics to help vehicle users get better LOS coverage, which is crucial for positioning and other applications.

\subsubsection{Coverage Benefit Harnessed by RISs}
We represent the total user population as the sum of the handset user Poisson point process and the vehicle user Cox point process. Then, we employ the Palm distribution of each user point process to feature typical users associated with each Palm distribution. To evaluate the connectivity improved by RISs, we introduce the LOS coverage metric, designed to illustrate the likelihood that a typical user is in LOS with respect to a base station or an RIS, provided that its SNR exceeds a certain threshold. We first determine whether a typical user is in LOS with respect to either base stations or RISs. Then, taking advantage of the probability generating functional of the point processes, we derive the coverage probability of the typical user as a closed-form expression. The derived formula provides the network performance as a function of its geometric parameters, such as the path loss exponent ratio ($\rho$), the blockage parameters ($\eta_1$ and $\eta_2$), the density of RISs ($\mu$), the density of base stations ($\lambda$), the density of vehicle users ($\nu$), and the density of handset users ($\lambda_2$). The obtained functional expression provides a straightforward means to individually assess the impact of geometric parameters on the network performance.

\subsubsection{Practical Network Design Insights for Various Deployment Scenarios}
The derived coverage probability of the network allows network operators and network designers to predict and configure the performance behavior of the RIS-enabled cellular system without the need for computationally intensive system-level simulations. Specifically, by leveraging the closed-form expression for the coverage probability, various network variables—such as the RIS density or the path loss exponent of road areas—can be individually or jointly tested, explicitly providing their impacts on network performance. To elaborate, we conduct a multi-variable analysis, providing high-level insights into the network performance behavior by presenting various numerical results. Examining the variation of the coverage probability with the number of RISs, we found that the performance of the RIS-assisted network is dictated by the local geometry. Furthermore, we observe that as the path loss or blockage difference between urban and road areas increases, the benefit of employing RISs in road areas is pronounced. This behavior justifies our design of deploying RISs in proximity to roads, requiring one to consider the network's local geometries and their differences for RIS deployment. We note that the gain grows as the number of vehicle users increases because, even though the reflected signals from RISs reach both handset and vehicle users, vehicle users at greater distances are more likely to receive the reflected LOS signal due to the relatively low attenuation and low blockage probability for links on roads. Throughout our various theoretical results, we show that the proposed RIS deployment strategy effectively harnesses the spatial correlation and local geometry within cellular networks with roads.
\section{System Model}
In this work, we exploit an approach to enhancing the large-scale performance of a RIS-enabled cellular network with vehicle users. To this end, this section builds an analytical framework modeling the locations of the network's spatial elements and emphasizing signal propagation dictated by the locations of links. Then, we define the coverage probability used in this work. 

\subsection{Spatial Model}
To model the locations of all base stations in dense urban areas, we use a Poisson point process of intensity $\lambda$ on $\bR^2.$  The base station Poisson point process is denoted by \[\Phi = \sum_{i\in\bN}\delta_{X_i},\] 
where $\delta_{x}$ is a Dirac-delta measure at $x\in\bR^2.$

This paper concerns a scenario where network users are (i) handset users randomly distributed on the urban areas and (ii) vehicle users clustered on the road areas. Therefore, we have 
\begin{align}
	\chi = \chi_1+\chi_2=\sum_{i}\delta_{V_{i}}+\sum_{i}\delta_{R_i}, 
\end{align}
where $\chi$ denotes the total user point process, $\chi_{1}$ denotes the vehicle user point process, and $\chi_2$ denotes the handset user point process. 

First, to model the locations of handset users, we use a Poisson point process of intensity $\lambda_2$. Then, to model the locations of spatially clustered vehicle users around the roads, we use a Poisson line Cox point process \cite{morlot2012population,8419219}. The Cox point process was extensively used and deemed necessary in the analysis of vehicular networks and cellular networks \cite{8340239,8419219,9204398}. Specifically, a Cox point process is given by the collection of the Poisson point processes, conditionally on a Poisson line process \cite{chiu2013stochastic,baccelli2010stochastic,baccelli2010stochasticvol2}. We denote by $\Phi_l$ the Poisson line process whose lines are created by the points of a Poisson point process  $\Xi$ of intensity $\lambda_{l}/\pi$ on a cylinder set $\cR = \bR\times [0,\pi].$ Then, for each point $(r,\theta)$, the variable $r$ gives the distance from the origin to the line $l(r,\theta)$. The variable $\theta$ gives the angle that the line $l(r,\theta)$ makes with the $x$-axis. The locations of the vehicle user are modeled as the Poisson point processes of intensity $\nu$ conditionally on those lines. 

\par To provide strong reflected signals and/or less severe blockage to vehicle users, we strategically deploy RISs on road areas so that RISs exploit the unique local network geometries of road areas. Since RISs and vehicle users must share the same road network, the Poisson line process $\Phi_l$ that gives rise to the RIS Cox point process must be the same Poisson line process $\Phi_l$ that gives rise to the vehicle user Cox point process $\chi_1$. Therefore, this model captures the spatial correlation of not only between vehicle users but also between vehicle users and RISs. The RIS Cox point process is given by 
\[\Psi = \sum_{k\in\bZ}\phi_k = \sum_{k\in\bZ}\sum_{m\in\bZ} \delta_{T_{m;k}},\]
where $\phi_k$ is the linearly distributed RIS Poisson point process of intensity $\mu$ on the line $l_{k }= l(r_k,\theta_k)$. Here, $T_{m;k}$ denotes the $m$-th RIS on the $k$-th line. Fig. \ref{fig:modelpicture01} shows the proposed model where road angles are isotropic whereas Fig. \ref{fig:modelpicture01_man} shows the proposed model where road angles are restricted to either  $0 \degree $  or $ 90 \degree$. 

\begin{remark}
We assume RISs are distributed as a random Poisson point process to account for practical limitations, such as the lack of feasible space or locations to deploy RISs uniformly. In such cases, network operators may deploy RISs in an ad-hoc fashion under various criteria. One of these criteria is to deploy RISs probabilistically, ensuring a certain average number of RISs of a given area. The Poisson point process is a general model that represents such a deployment scenario.
	
Compared to a regular distribution scenario where inter-RIS distances is fixed as a constant, our model assumes RISs are distributed as a Poisson point process, which conditionally realizes the uniform distribution of RISs on a finite segment given the number of RISs in that segment.
	
For instance, if the linear density of RISs per kilometer is $5$, the number of RISs follows a Poisson distribution with a mean of $5$ for a road segment of length $1$ kilometer, resulting in $5$ RISs on average in that segment. The average distance between RISs is $200$ meters. For regularly-distributed RISs with the same density, there are also five RISs on average per kilometer, and the average distance between RISs is similarly $200$ meters. Consequently, on average, both regularly-distributed and Poisson-distributed RISs exhibit the identical RIS layout.
\end{remark}
\begin{example}
It is worthwhile noting the difference between regularly-distributed RISs and Poisson-distributed RISs. For regularly-distributed RISs, the distance from a user on the road to its closest RIS follows a uniform distribution between 0 and $100$ meters. In contrast, for Poisson-distributed RISs, the distance from a user on the road to its closest RIS follows an exponential distribution with a mean of $100$ meters. This difference in distance distribution is one of the key motivations for using the Poisson point process in our model, as it provides a more flexible and realistic representation of potential RIS deployments in practical scenarios. Regularly distributed RISs will generally provide an overestimation of the connectivity probability, as the typical vehicle user always finds its RISs within a certain distance. The performance analysis based on regularly distributed RISs is outside the scope of this paper and is left for future work. 
\end{example}

\subsection{Signal Propagation and Blockage}
\begin{figure}
	\centering
	\includegraphics[width=1\linewidth]{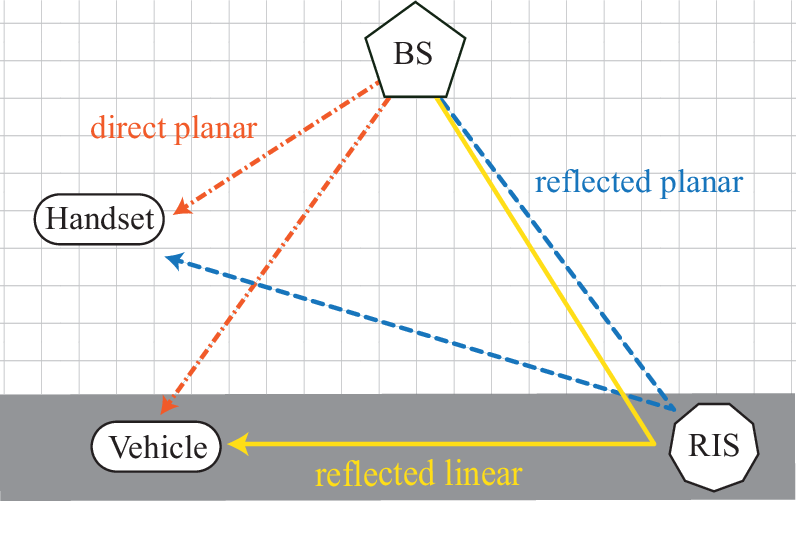}
	\caption{In this paper, only RIS-to-vehicle links are on roads and they are referred to as linear links having path loss exponent $\alpha_1$ and blockage parameter $\eta_1$. All other links are planar links having path loss exponent $\alpha_2$ and blockage parameter $\eta_2.$}
	\label{fig:distinctpl}
\end{figure}
We employ a distance-based power-law path loss model, mainly motivated by the widely-accepted 3GPP standard \cite{38901}.  Specifically, we assume that the received signal power of the receiver at a distance $ d $ from a transmitter is 
\begin{align}
	\begin{cases}
 		 {\frac{p_tG_tG_r(\lambda_c/4)^2}{L_pL_r}}d^{-\alpha_1} &\text{ for links in road areas},\\
		{\frac{p_tG_tG_r(\lambda_c/4)^2}{L_pL_r}} d^{-\alpha_2} &\text{ for links in urban areas},
	\end{cases}
\end{align}
where $p_{t}$ represents the transmit power, while $G_t$ and $G_r$ denote the antenna gains available at the transmitter and receiver, respectively. Additionally, $L_p$ is a constant that relates transmit and receive signal power, and $L_r$ accounts for implementation loss. $\lambda_c$ is the wavelength of the signal. 

We use different path loss exponents for road and urban areas with $\alpha_2 > \alpha_{1} > 2$. These path loss exponents define the level of path loss, and their specific values are carefully chosen to match the distinctive propagation patterns of signals over areas with different local geometries in urban environments \cite{38901}. For instance, signals in urban areas experience significant attenuation due to the presence of numerous large-sized obstacles or reflectors, such as buildings. In contrast, signals traveling over roads may suffer less severe attenuation compared to urban areas because there are typically fewer small-sized obstacles, such as vehicles. To distinguish between the path loss exponents applied to different signal propagation environments, we also refer to the link in urban areas as `planar link' and the link in road areas as `linear link'. In Fig. \ref{fig:distinctpl}, the link from RIS to vehicle is the linear link. 

\par In a similar manner, we utilize a distance-based independent blockage model, building upon the random blockage model in \cite{38901,6290250,9174910,9868205}. Specifically, we define $\bP(B_i)$ as the blockage probability of link $i$, characterized the following probability:
\begin{align}			
	\bP(B_i)=
	\begin{cases}
 	1-\exp(-d_i/\eta_1)&\text{for linear links},\\ 
		1-\exp(-d_i/\eta_2) &\text{for planar links}, \label{eq:blockage}
	\end{cases}
\end{align}
where $d_{i}$ represents the distance of link $i$,  $\eta_1$ is the blockage parameter associated with linear links, and $\eta_2$ is the blockage parameter associated with planar links. In general, links in urban areas are more prone to blockages compared to links traveling over roads \cite{38901}, and thus we assume that $\eta_2$ is greater than $\eta_1$. Under a general setting, the blockage parameters $\eta_1$ and $\eta_2$ roughly correspond to the average LOS distances---average distance to the first obstacle---of links on road and on urban areas, respectively \cite{9759490,10024366}. 

\begin{remark}
The LOS probability must depend on the distribution of obstacles in the environment. In this paper, for both simulation and analysis based on stochastic geometry, we employ the widely used random blockage model based on 3GPP\cite{38901}, where the LOS probability decreases exponentially with the distance of the link between transmitters and receivers, and the rate of decrease is captured by the blockage coefficient that depends on the local surrounding of the link. For the blockage and attenuation parameters, we present a simple expression by leveraging the numerically-proven LOS probability and path loss expressions provided in \cite[Section 7.4.1]{38901}. Specifically, the hyper-parameters for blockage or path loss for various links in this paper, such as BS-to-user, BS-to-RIS, and RIS-to-vehicle links, are directly obtained from the blockage probability and path loss expressions of UMa channel as specified by \cite[Section 6.2]{37885}. Note that the exponential characterization of the LOS probability appears in the 3GPP document and in numerous papers including \cite{6290250,9174910,9868205} since it reflects the essential geometric fact that the likelihood of being obstructed by an obstacle increases exponentially as the distance of the link increases.
\end{remark}

\subsection{RIS Operation}\label{S:RIS}
In our proposed network model, RISs are strategically positioned along roads to provide reflected signals to users. Specifically, this paper concentrates on assessing the large-scale geometric impact of these reflected signals by examining changes in users' LOS connectivity and signal coverage. 
To this end, we simplify the operation of RISs, assuming that an RIS would provide reflected signals to any nearby users if it receives an unobstructed LOS signal from a base station, by adjusting the RIS phase shifts.

This simplified operation of RISs allows us to explore the theoretical potential of RIS-enabled networks while focusing on network's geometric constraints. It avoids introducing any hardware limitations, which could complicate the theoretical and fundamental analysis of the network connectivity.
\begin{remark}
In this paper, the traditional cell boundary (such as the Voronoi boundary) does not apply to the service boundary that determines which base station provides the LOS signal to the users. Instead, we actively employ the concept of the 'LOS' area, where the 'viable' (above some threshold) LOS signal from the base station will be received. Depending on the blockage and the relative locations of base stations and RISs, an RIS is enabled or not, based on whether it is within the LOS area of a base station (or base stations). Only those enabled RISs reflect signals from base stations to their surrounding users, providing LOS signals to the terminal users. With these reflected signals, users will have a higher chance of receiving LOS signals with enough signal power to facilitate the necessary applications. This work evaluates the benefits of the proposed RIS system, analyzing them through integral expressions that explicitly illustrate various key geometric and propagation parameters.
	
	Our system assumes the proposed RIS operation by weakly associating RISs and users with their closest base stations, based on the traditional cell boundary. It is worth noting that the list of RISs within the LOS area of base stations might be maintained by those base stations (or the core network), provided that base stations and roadside RISs are stationary. Once the core network detects users with very low throughputs (or users report high positioning errors to the system), the core network initiates the procedure of providing an alternative LOS signal with the help of RISs. The network operator may first utilizes its listed LOS RISs to find the best RIS that can better serve the user. Since vehicle users are mostly located on roads, and roadside RISs reflect the signals from stationary base station toward the road infrastructure, the number of reflection coefficients viable at RISs is limited to certain value and the reflection coefficient needs to be found through an exhaustive search coordinated by the core network. Once the network user’s throughput rises (or its positioning error drops), the user is expected to receive the reflected LOS signal through an RIS from a base station.
\end{remark}

\begin{figure}
	\centering
	\includegraphics[width=.85\linewidth]{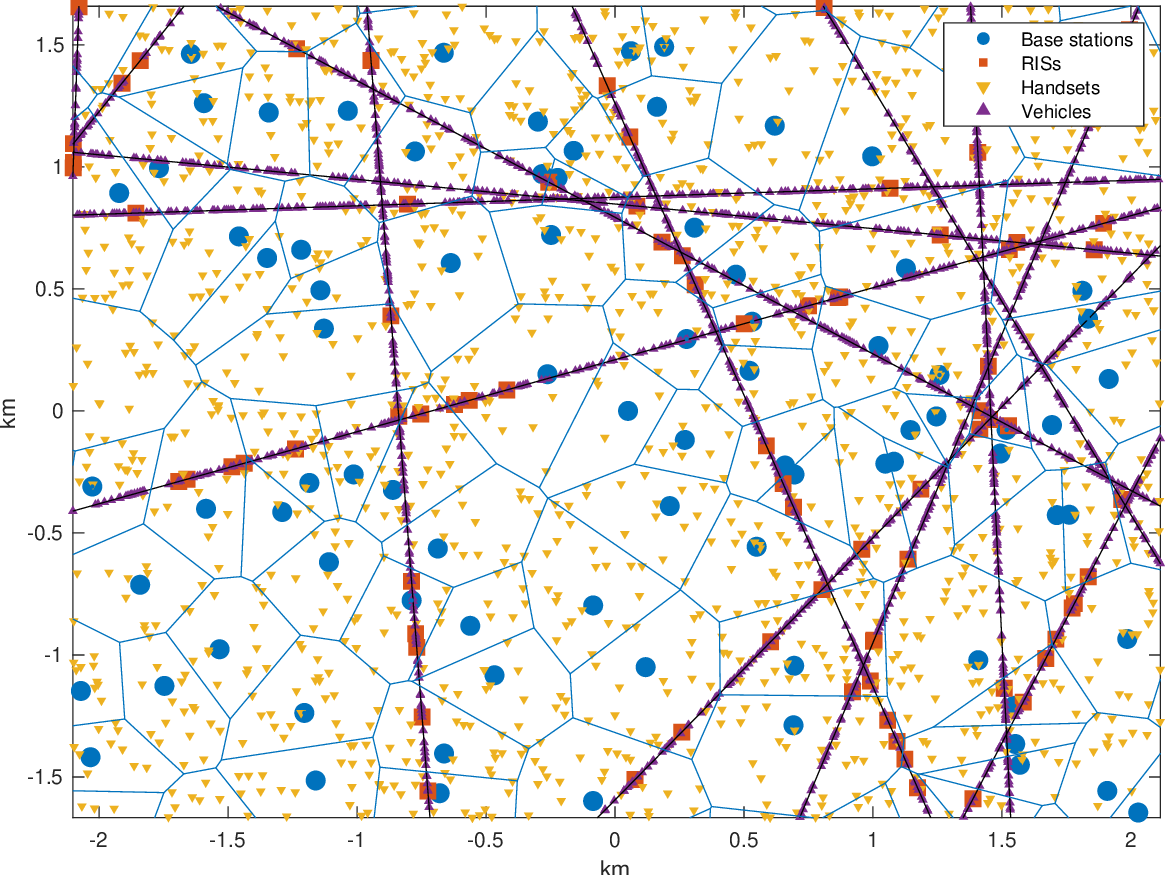}
	\caption{Base stations and handset users are Poisson point processes of densities $3/\text{km}^2$ and $100/\text{km}^2$, respectively. RISs and vehicles users are Cox point processes with densities $\lambda_l\mu=10/\text{km}^2$ and $\lambda_l\nu = 100/\text{km}^2,$ respectively.}
	\label{fig:modelpicture01}
\end{figure}
\begin{figure}
	\centering
	\includegraphics[width=.85\linewidth]{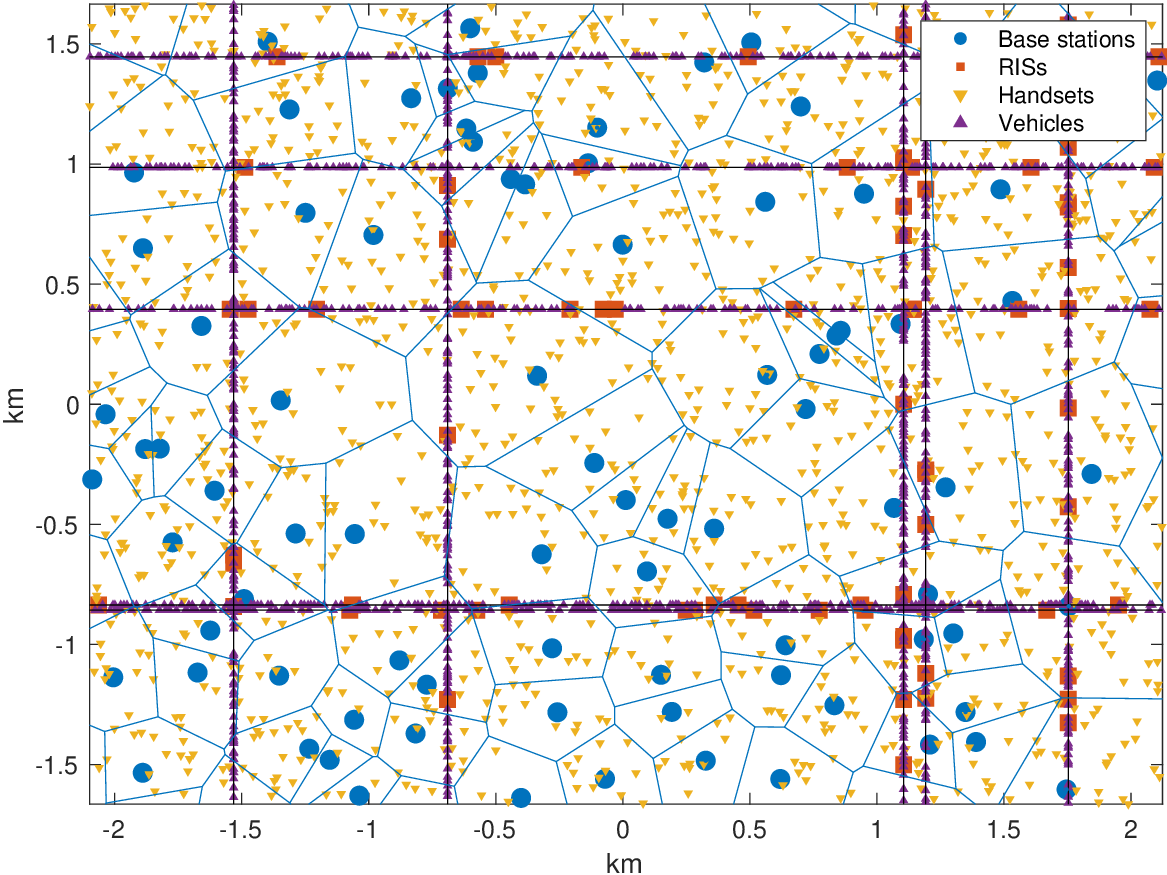}
	\caption{Base stations and handset users are Poisson point processes of density $3/\text{km}^2$ and $100/\text{km}^2$, respectively. RISs and vehicles users are Cox point processes of dentieis $\lambda_l\mu=6/\text{km}^2$ and $\lambda_l\nu = 100/\text{km}^2,$ respectively.}
	\label{fig:modelpicture01_man}
\end{figure}

 \subsection{Performance Metric: Coverage Probability}
In this paper, we aim to analyze the theoretical limit of RIS-assisted connectivity. To this end, we ignore any impact of other-cell interference. Then, a network user is defined as being in coverage if it receives an unobstructed LOS signal either from a base station or an RIS, and the average SNR of such a LOS signal exceeds a given threshold $\tau$. 

Since the user point process is stationary, we consider a Palm distribution of the user point process to feature a typical user at the origin.  The Palm distribution of the user point process is given by the summation of the Palm distributions of each user point process\cite{baccelli2010stochastic}.  As a result, the coverage probability of the typical user is given by 
\begin{align}
	\bP_{\chi}^0(\textrm{cov}) 
	&=\frac{\lambda_{l}\nu}{\lambda_{l}\nu+\lambda_2}\bP_{\chi_{1}}^0(\text{cov})+ \frac{\lambda_2}{\lambda_{l}\nu+\lambda_2} \bP_{\chi_2}^0(\text{cov})\label{eq:total},
\end{align}
where $\lambda_{l}\nu$ and $\lambda_2$ are the average numbers of the vehicle users and handset users in a unit area, respectively. In Eq. \eqref{eq:total}, $\bP_{\chi_1}^0(\textnormal{cov})$ is referred to the coverage probability of the typical vehicle user and $\bP_{\chi_2}^0(\textnormal{cov})$ is referred to as the coverage probability of the typical handset user. For both cases, the event $\{\textnormal{cov}\}$ occurs if 
\begin{equation}
	\{\text{the typical user has an LOS TX w/ {avg SNR} $>\tau$ }\}\nnb.
\end{equation}
This event is further divided into two cases where (i) the typical user get signal directly from base station or (ii) indirectly from an RIS.  First, 
if the typical user directly receives signals from a base station, the average SNR is given by



\begin{equation}
	\frac{p_tG_tG_r (\lambda_{c}/4)^2  }{L_pL_rN_0B_WF} d^{-\alpha_2}\label{direct},
\end{equation}
where $N_0=-174$ dBm/Hz, $B_W$ is the bandwidth, and $F$ is the noise figure. 

On the other hand, with RISs, the average SNR of the typical user is given by 
\begin{equation}
	\frac{p_tG_t G_r (\lambda_{s}/4)^2  }{L_pL_rN_0B_WF} N_r^2d_{1}^{-\alpha_2}d_2^{-\alpha} \label{reflected},
\end{equation}
where $N_r $ is the number of RIS elements. The distance from the base station to the RIS  is denoted by $d_1$ and the distance from the RIS to the typical user is denoted by $d_2$.  In Eq.  \eqref{reflected}, if the typical user is vehicle user, the exponent associated with $d_2$ is given by $\alpha=\alpha_1$; otherwise, the exponent holds $\alpha=\alpha_2.$ 

For a simpler notation, this paper defines $\gamma$ as follows: 
\begin{equation}
	\gamma= (p_tG_t G_r(\lambda_c/4)^2)/(L_pL_rN_0B_{W} F). \label{maxpl}
\end{equation}

It is worth noting that we consider the average SNR of the signal and then make the coverage probability our performance metric to assess the limitations of RIS-assisted cellular networks. The methodology and performance metric highlight the effects of geometric dependence and blockage. Since we leverage the stationary point process for modeling users, our derived coverage probability of the typical user is the coverage averaged across all the users in the network.

\begin{remark}
In this paper, we ignore the interference created by other base stations or RISs because, as the performance metric we chose to examine shows, we focus on the fundamental limit of the network connectivity—specifically, the maximum chance of coverage that users may achieve when served by both spatially independent base stations and spatially correlated RISs. Due to interference and hardware limitations discussed in Section \ref{S:RIS}, the real-world coverage probability must be lower than the one assessed in this paper. Nevertheless, this performance metric is a guideline to the maximum achievable performance of RISs, whose effectiveness is limited by network geometry.

Besides, we have chosen to simplify the small-scale fading. The reason for this simplification is to clearly emphasize the geometric limitations and benefits of spatially correlated RISs and vehicle users while allowing us to mathematically derive the interactions between network parameters based on stochastic geometry. We acknowledge that small-scale fading causes the received signal power to vary around its mean, which could obfuscate the impact of the geometric characteristics we aim to highlight. Our objective is to provide a clear understanding of how the spatial correlation between RISs and network users---as well as the favorable propagation conditions on various area---contribute to network performance. 
\end{remark}

Throughout this paper, we denote by $\bP(X)$ as the probability of random variable or event $X$.  Suppose $\chi$ is a stationary point process. Then,  $\bP_{\chi}^0(X)$ is the probability of  a random variable or event $X$ under the Palm distribution of $\chi$; this means the probability of $X$ conditionally on a typical point of $\chi$, located at the origin.  Similarly, $\bE[Y]$ is the expectation of a random variable or event $Y$. Suppose $\chi$ a stationary point process. Then,   $\bE_{\chi}^0(Y)$ is the expectation of a random variable or event $Y$ under the Palm distribution of $\chi$; this means the expectation of $Y$ conditionally on a typical point of $\chi$, located at the origin. Suppose $\chi_1 $ and $\chi_2$ stationary point processes. $\bP_{\chi_1+\chi_2}^0(Z)$ is the probability of $Z$ under the Palm distribution of $\chi_1+\chi_2$. See \cite{baccelli2010stochastic,haenggi2012stochastic} regarding the Palm distribution of stationary point processes.

\begin{table}
	\caption{Network Parameters}\label{Table:1}
	\centering	
	\begin{tabular}{|c|l|}
		\hline
		Var & Description \\
		\hline
		$\Phi_l$ & Road Poisson line process $\lambda_l/\pi.$\\ 
		\hline 
		$\Phi$ &  Base stations Poisson point process (density:$\lambda $) \\
		\hline 
		$\Psi$ & RIS Cox point process $(\lambda_l\mu)$\\
		\hline
		$\chi$& Total user point process ($\chi = \chi_1+\chi_2$)\\
		\hline
		$\chi_{1}$ & Vehicle user Cox point process (density:$\lambda_l\nu$)\\
		\hline 
		$\chi_2$ & Handset user Poisson point process (density:$\lambda_2$) \\ \hline 
		${\alpha_2}$ & Path loss exponent for planar link \\ \hline 
		${\alpha_1}$ & Path loss exponent for linear link\\ \hline 		
		$\eta_2$ & urban areas blockage parameter \\ \hline 
		$\eta_1$ & road areas blockage parameter \\ \hline 
		$\tau$ & average SNR threshold \\ \hline
	\end{tabular}
\end{table}

\section{The Coverage Probability}
This section analyzes the coverage probabilities of the typical user without and with RISs, respectively. By expressing the coverage probabilities as functions of network variables, we learn the fundamental limit of the large-scale connectivity of the RIS-assisted cellular network as an explicit function of geometric parameters including the average number of RISs in a given area. 
\subsection{Coverage Probability without RIS}
{In this subsection, we first evaluate the coverage probability of the network without any RISs in the system. The coverage probability will be given by a simple closed form formula and it will be later used to demonstrate the benefit of the proposed strategic RIS deployment on roadsides.}
\begin{lemma}\label{Theorem:1}
	Without RISs, the coverage probability of the typical user is given by 
	\begin{align}
	1- \exp\left({\left.-2\pi\lambda{\int}_{0}^{c_0}r\exp\left({-r/\eta_2}\right)\diff r\right.}\right), \label{eq:theorem1}
	\end{align}
	where we have 
	\begin{equation}
		c_0={\left({\gamma}/{\tau}\right)^{{1}/{\alpha_2}}},
	\end{equation} and $\gamma /\tau$  is the maximum tolerable path loss. 
\end{lemma}
\begin{IEEEproof}
See Appendix \ref{A:1}.
\end{IEEEproof}

The coverage formula demonstrates that the coverage probability increases when (i) the average density of base stations denoted by $\lambda$ increases, (ii) the blockage parameter for urban links denoted by $\eta_2$ decreases, or (iii) the path loss exponent for urban links denoted by $\alpha_{2}$ decreases. Similarly, the outage probability, defined as $1-\bP(\text{cov})$, decreases exponentially with the base station density. 

Since the coverage probability is derived for the typical user at the origin under the Palm distribution of the user point process,  it characterizes the coverage probability of all users, i.e., the coverage spatially averaged across all the users in the network. Therefore, the derived metric in fact reveals the large-scale behavior of the LOS coverage without RISs. 

 \begin{figure}
 	\centering
 	\includegraphics[width=1\linewidth]{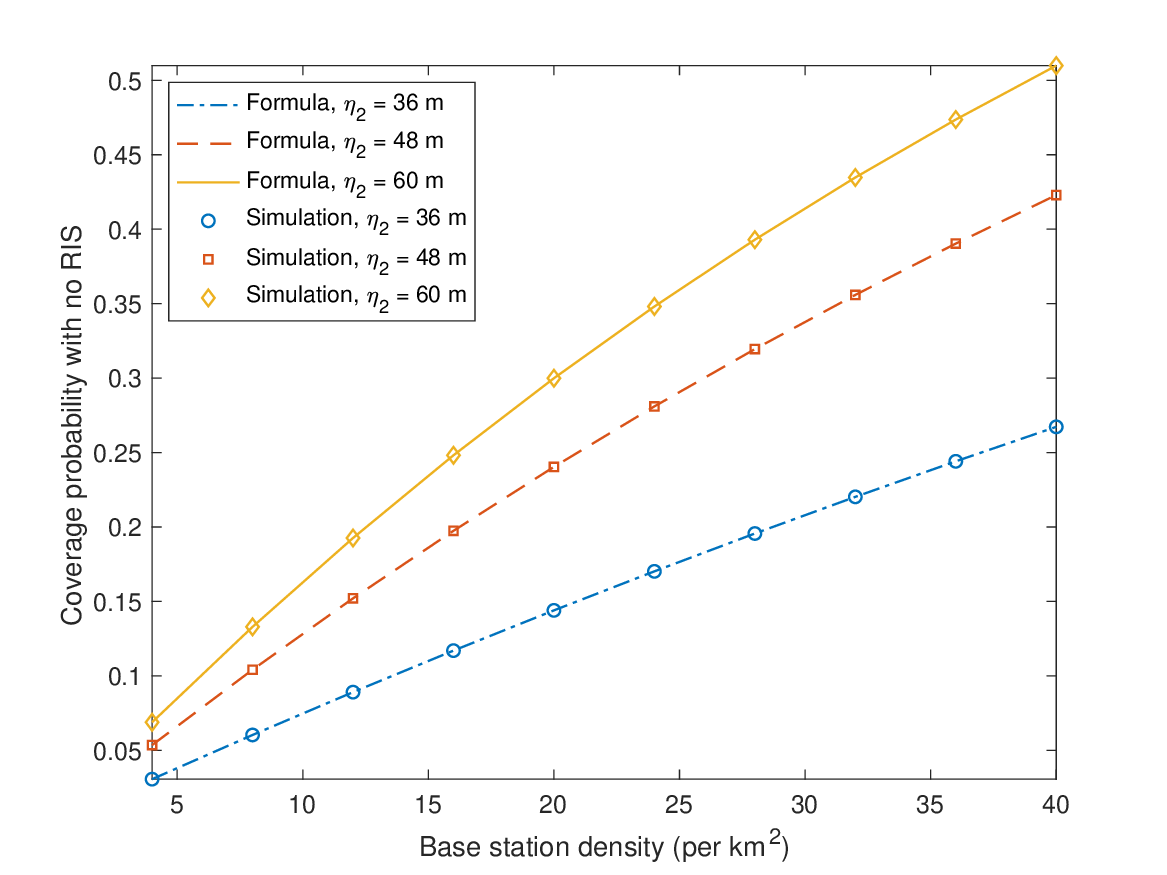}
 	\caption{{The coverage probability of the typical user without RISs. We use $\gamma=83 $ dB.}}
 	\label{fig:losworis_monte}
 \end{figure}
\begin{figure}
	\centering
	\includegraphics[width=1\linewidth]{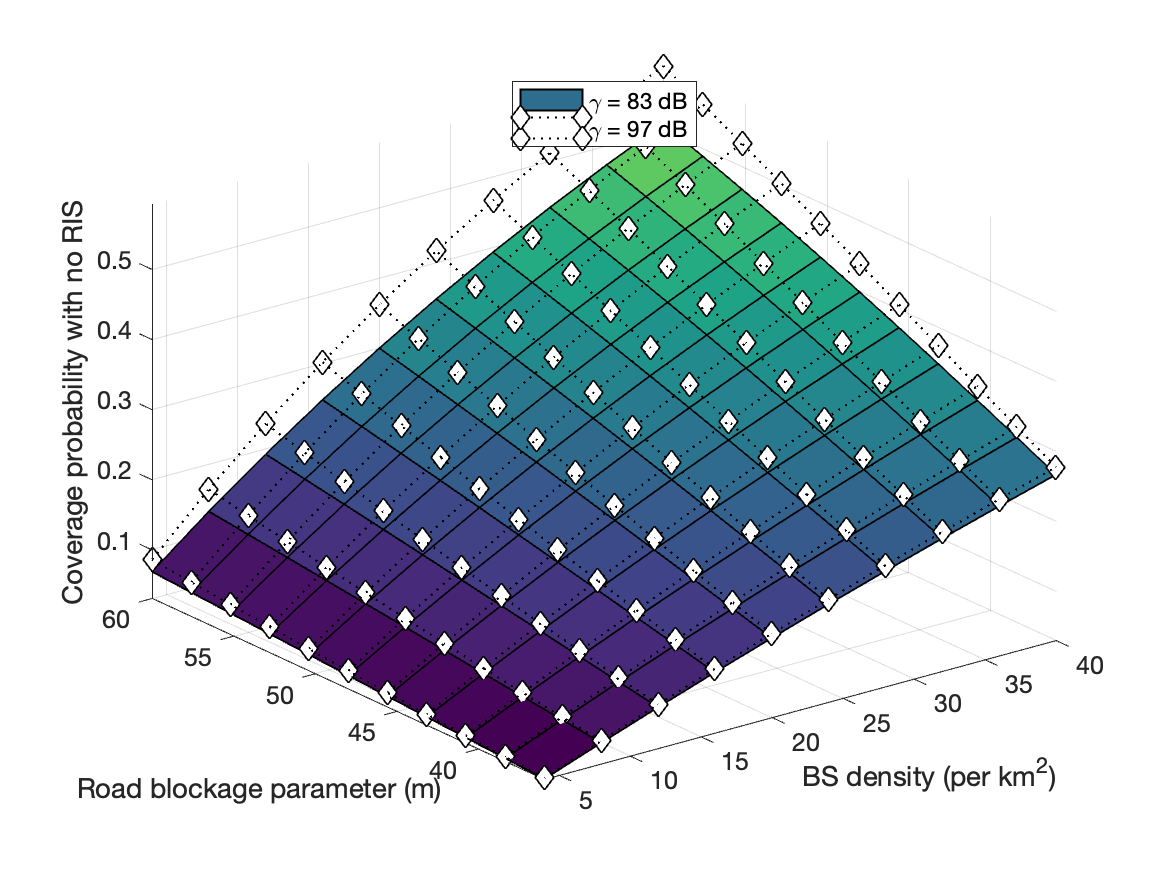}
	\caption{The coverage probability with no RIS. $\gamma=83$ dB and $\gamma=97$ dB.}
	\label{fig:fig1noriscoverage}
\end{figure}


The simulation results in Fig. \ref{fig:losworis_monte} validate the accuracy of the derived formula.  In 
Figs. \ref{fig:losworis_monte}  and \ref{fig:fig1noriscoverage}, we use $p_t=23$ dBm, $G_t=5$ dB, $G_r=5$ dB, $f_c=6$ GHz, $N_0=-174$ dBm, $B_W=10$ MHz, $F=6$ dB, $N_r^2=36$, $\tau=0$ dB, and $\alpha_2=3.7$. The base station density varies from $4/\text{km}^2$ to $40/\text{km}^2$ to show the coverage probability for various $\lambda$.  

In Fig. \ref{fig:fig1noriscoverage}, we compare different scenarios $\gamma=83$ dB and $\gamma = 97$ dB, respectively. The coverage probability of $\gamma=97$ dB is greater than that of $\gamma=83$ dB. Nevertheless, the additional $10$ dB to $\gamma$ does not sharply enhance the coverage probability of the network.  



\begin{figure*}
	\begin{align}
		\bP_{\chi_{1}}^0(\text{cov}) = &1-\exp\left(-2\pi\lambda\int_{0}^{c_0}re^{-\frac{r}{\eta_{2}}}\diff r-\mu\int_0^{c_1}\left\{1-f_{1}(t)\right\}\diff t -{\lambda_l}\!\int_0^{c_2}1-e^{-2\mu\int_0^{\sqrt{c_2^2-u^2}}\{1-f_{2}\left(\sqrt{u^2+t^2}\right)\}\diff t }\diff u \right)\label{eq:linearcov},\\ 
		\bP_{\chi_{2}}^0(\text{cov})=& 1-\exp\left(-2\pi\lambda\int_{0}^{c_0}re^{-\frac{r}{\eta_{2}}}\diff r-{\lambda_l}\!\int_0^{c_2}1-e^{-2\mu\int_0^{\sqrt{c_2^2-u^2}}\{1-f_{2}\left(\sqrt{u^2+t^2}\right)\}\diff t }\diff u \right)\label{eq:planarcov}.
	\end{align}
	\rule{\linewidth}{0.2mm}
\end{figure*}

\subsection{Coverage Probability with RIS}

By exploiting distinctive network geometries manifested in planar and linear links, RISs can reflect signals from base stations, allowing network users in various locations to bypass the blockage of direct signals or excessive attenuation. In this subsection, we analyze the coverage probability of the typical user in the network with RISs.

We will see that the favorable geometric characteristics of links over roads, namely its small blockage parameter or small path loss exponent are exploited by placing RISs on roads. This eventually enhances the coverage probability of the network. We provide the analytical result first. Then we give interpretations and observations. 


\begin{theorem}\label{theorem:2}
	With RISs, the coverage probability is given by 
	\begin{align}
\frac{\lambda_l\nu}{\lambda_l\nu+\lambda_2} \bP_{\chi_{1}}^0(\textnormal{cov}) + \frac{\lambda_2}{\lambda_l\nu+\lambda_2} \bP_{\chi_{2}}^0(\textnormal{cov}),\label{eq:theorem2}
	\end{align}
	where $\bP_{\chi_{1}}(\textnormal{cov})$ and $\bP_{\chi_{2}}(\textnormal{cov})$ are given by Eqs. \eqref{eq:linearcov}  and \eqref{eq:planarcov}, respectively. We define 
$c_0=(\gamma/\tau)^{1/\alpha_2}, $ $c_1= (\gamma N_r^2/\tau)^{1/\alpha_1}$, and $c_2=(\gamma N_r^2/\tau)^{1/\alpha_2}. $ We also define the functions as follows: 
	\begin{align}
			f_{1}(t) &= (1-e^{-{t}/{\eta_1}})\exp\left(-2\pi\lambda\int_{0}^{{c_2}/{t^{\rho}}} r e^{-\frac{r}{\eta_2}}\diff r\right),\\
		f_{2}(t) &= (1-e^{-t/\eta_2})
	\exp\left(\left.-2\pi\lambda\int_{0}^{{{c_2}}/t}  r e^{-r/\eta_2}\diff r\right.\right).
	\end{align}
\end{theorem}

\begin{IEEEproof}
See Appendix \ref{A:2}.
\end{IEEEproof}

Here are a few observations we derive out of Lemma \ref{Theorem:1} and Theorem \ref{theorem:2}. First, we obtain the coverage probability of the network with RISs as an explicit function of all network variables $N_r^2, \alpha_1,\alpha_2,\eta_1,\eta_2,\lambda,\lambda_2,\lambda_l,\mu$, and $\nu.$ Therefore, the behavior of the coverage probability w.r.t. these parameters is directly identified and even accurately predicted by just leveraging the derived formula. As other existing work \cite{6042301} leveraging stochastic geometry, assessing the large-scale network performance without extensive and numerous system-level simulations results in significant reduction in computation cost. Eventually the present analysis will help us to optimize design of the network in practice.

Furthermore, Theorem \ref{theorem:2} represents the coverage probability as a function of the average numbers of the handset users and vehicle users. For Eq. \eqref{eq:theorem2},  
$\bP_{\chi_{1}}(\textnormal{cov})$ is the coverage probability of the typical vehicle user. To evaluate it, we feature the typical vehicle user at the origin. On the other hand, $\bP_{\chi_{2}}(\textnormal{cov})$ is the coverage probability of the typical handset user. To evaluate it, we feature the typical handset user at the origin. Then, these individually computed two coverage probabilities are weighted by coefficients and then they are combined to determine the final network coverage probability. Each coefficient is determined by the relative density of the corresponding user to the total user density. For instance, if there are more handset users than vehicle users, in Eq. \eqref{eq:theorem2}, the second coefficient dominates the first coefficient. On the other hand, if there are more vehicle users than handset users, the first coefficient dominates the second coefficient. 

From individually examining the expressions Eq. \eqref{eq:linearcov} and \eqref{eq:planarcov}, we define of the outage probability of the typical user to obtain the following expression. 
\begin{equation} \label{111}
	\frac{1-\bP_{\chi_{1}}^0(\textnormal{cov})}{1-\bP_{\chi_{2}}^0(\textnormal{cov})}=\frac{\bP_{\chi_{1}}^0(\textnormal{outage})}{\bP_{\chi_{2}}^0(\textnormal{outage})} = e^{-\mu\int_{0}^{c_1}\{1-f_1(t)\}\diff t } .
\end{equation}
The right hand side of Eq. \eqref{111} is always less than or equal to one. This gives us an important following observation: 
\begin{align}
	\bP_{\chi_1}^0(\text{cov})  \geq \bP_{\chi_2}^0(\text{cov}).
\end{align} 
Namely, the typical vehicle user has a greater coverage probability than the typical handset user thanks to the RISs deployed on road areas.In Eq. \eqref{111}, the equality holds only when $ \mu=0 $, namely there are no RISs. In other words, by displacing RISs, the vehicle users have a better coverage probability than the handset users.

By examining the formulas in Lemma \ref{Theorem:1} and Theorem \ref{theorem:2}, our analysis demonstrates that the coverage probability of a typical user assisted by RISs is higher than that of a typical user without RISs, assuming all other network parameters remain constant. This is based on the concept of stochastic dominance, where the CCDF of one random variable is always greater than the CCDF of another at any given value. By ensuring identical blockage and attenuation parameters in both scenarios, the improved coverage with RISs becomes evident. Our study mathematically quantifies the benefits of introducing RISs at a specific density, particularly when deployed on roadsides. It is very important to note that the coverage probability of the typical user is given by a function of all geometric parameters including $\alpha_1,\eta_1, \alpha_2, $ and $\eta_2.$

The right hand side of Eq. \eqref{111} is a function of path loss ratio $\rho = \alpha_1/\alpha_2$ and the expression exactly accounts for the impact of the difference in signal attenuation. For instance,  when the difference in the path loss exponents gets bigger, $f_1(t)$ becomes smaller accordingly and therefore the difference in the coverage probabilities grows larger.

The uniqueness of our performance analysis lies in the specific deployment of RISs near roadsides, which fundamentally differs from general RIS deployment scenarios. When RISs are not placed near roadsides, their network geometry is similar to that of base stations, resulting in the same blockage distribution and large-scale path loss for both. Although the coverage probability may increase with more RISs, the specific benefits of roadside deployment are not realized. Our study highlights the unique attenuation and blockage properties related to road infrastructure, which are not utilized in general RIS deployments. Empirical evidence \cite{37885,38901} shows that vehicle users are better served by roadside units than by base stations located far from road infrastructure.
	
By strategically deploying RISs near roads and using Cox point processes to model RISs and vehicle users simultaneously, we accurately capture the spatial correlation between them. This approach allows us to analyze different types of links among base stations, RISs, vehicle users, and other network users within a unified framework. The proximity of roadside RISs to vehicle users and the exploitation of reduced blockage and attenuation along roads demonstrate their potential to provide better coverage and improved network performance. In summary, our analysis uniquely accounts for the spatial dependence and distinct propagation characteristics of roadside deployments, offering valuable insights for scenarios with high densities of vehicle users and roadside infrastructure.
\begin{remark}
Recall that when RISs are not deployed near roadsides, they are effectively the same as base stations in terms of
	network geometry. Consequently, the propagation characteristics from these RISs to network users
	and from base stations to network users undergo the same blockage distribution and large-scale path
	loss. In such a setup, although the LOS coverage probability may increase due to the additional
	number of RISs, the unique benefits associated with roadside deployment cannot be realized.
\end{remark}

\begin{example}
Our current analysis assumes a full-disk reflection space for RISs, which represents the best possible coverage. However, our study can be extended to incorporate the impact of half-space reflection. In our model, each RIS's reflection space can be represented as a full disk. To account for a more realistic half-space reflection, one can modify the model to use a half-disk reflection space for each RIS, for instance. First, the orientation of these half-disks would then need to be determined; they could be randomly oriented or aligned parallel to the road. Similarly, an independent orientation approach could be employed, where each roadside RIS is assigned a randomly oriented half-disk, resulting in a Boolean model on the Cox point process. Then, by using techniques similar to those used in our current computation of SNR coverage probability, one can derive the LOS coverage probability while accounting for the half-space reflection constraint. A more detailed analysis based on the half-space reflection constraint is left for future work.
\end{example}

\section{Numerical Results}
This section shows the coverage probability of the network evaluated by Theorem \ref{theorem:2} and Monte Carlo simulations. In general, we use the network parameters provided in Table \ref{Table:2}. 

\begin{table}
	\centering
	\caption{Experiment Parameters}\label{Table:2}
	\begin{tabular}{|c|c|c|}
		\hline
		Variable & Parameters &  Values \\ 
		\hline 
		$\lambda$& Base station density & $40/\text{km}^2$ \\
		\hline 
		$\lambda_l$ & Road density & $2/\text{km}^2$\\
		\hline
		$\lambda_2$ & Handset user density &  $200/\text{km}^2$\\
		\hline 
		$\nu$&Vehicle user density &  $25/\text{km}$ \\
		\hline 
		$		\mu$ &RIS density &  $4$/km    \\ \hline 
		$\alpha_1$ & Road path loss exponent & $2.4$ \\ \hline 
		$\alpha_2$ & Urban path loss exponent & $3.7$  \\ 
		\hline
		$\eta_1$ & Road blockage parameter & $48 $ m  \\ \hline 
		$\eta_2$ & Urban blockage parameter  & $36$ m \\ \hline
		$\gamma$ & Maximum tolerable path loss & $97$ dB \\\hline 
		$p$ & Transmit power & 30 dBm \\\hline  
		$G_t$ & Transmit antenna gain  & 12 dB \\\hline 
		$G_r$ & Receive antenna gain  & 5 dB \\\hline 
		$N_o$ & Thermal noise per Hz & $-174 $ dBm/Hz \\\hline 
		$B_w$ & Bandwidth & $10$ MHz \\\hline 
		$F$ & Noise figure  & $6$ dB \\\hline 
		$N_r$ & RIS elements & $64$ \\\hline 
	\end{tabular}
\end{table}

\subsection{Impacts of SNR Thresholds and RIS elements}
\begin{figure}
	\centering
	\includegraphics[width=1\linewidth]{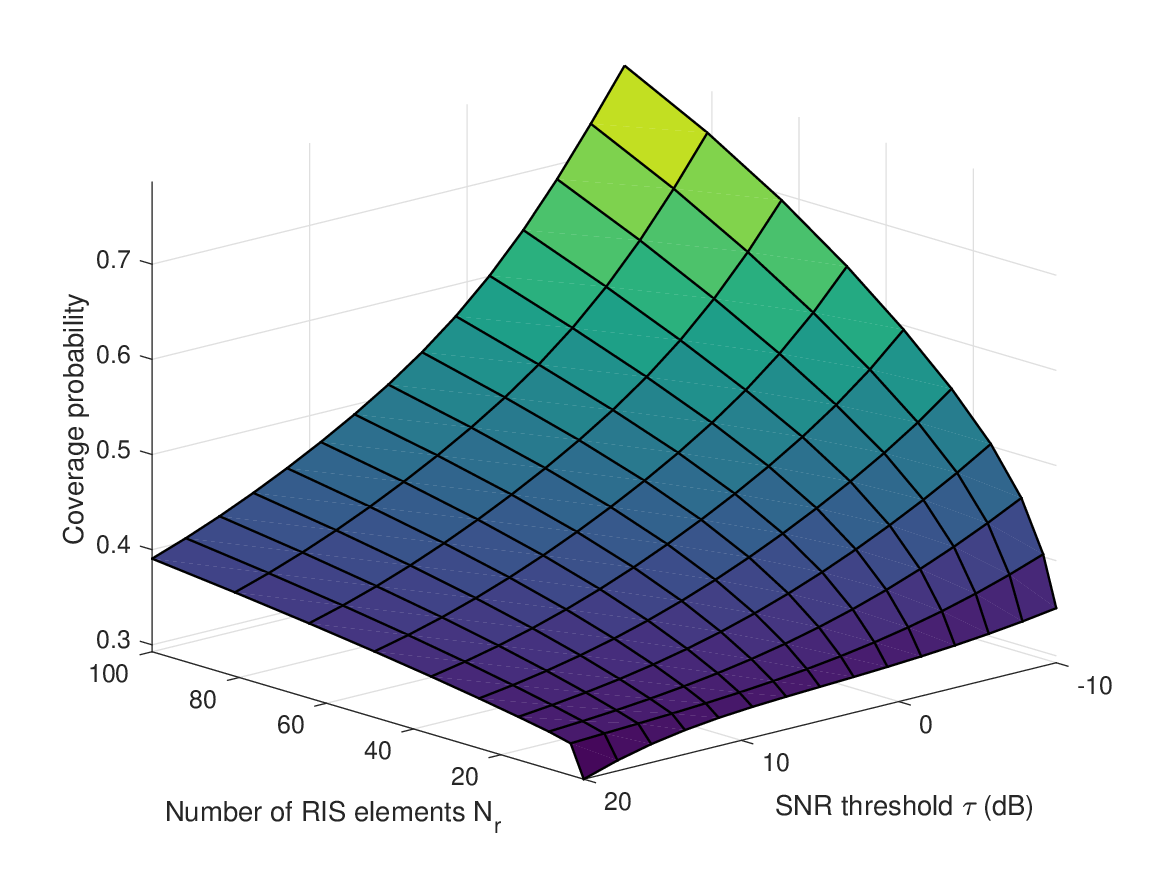}
	\caption{The coverage probability of the typical user. }
	\label{fig:riselements}
\end{figure}

\begin{figure}
	\centering
	\includegraphics[width=1\linewidth]{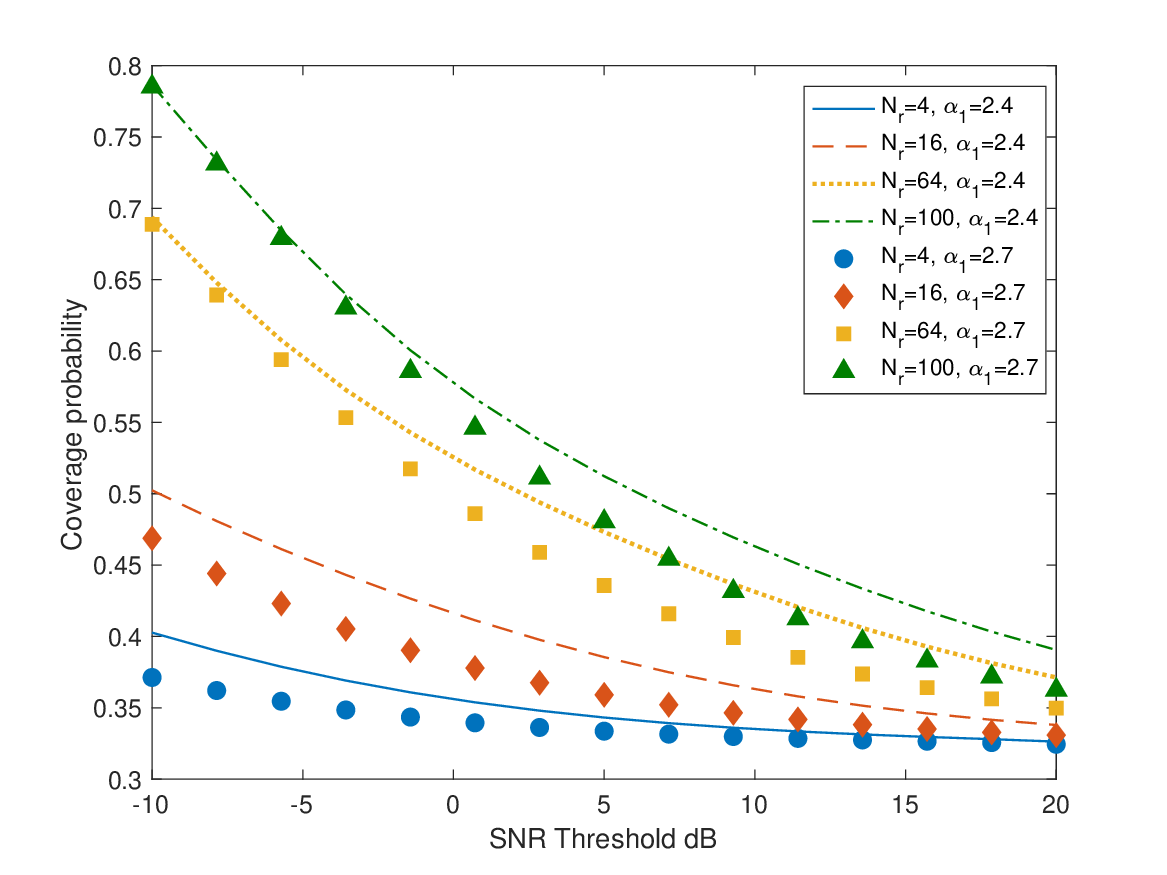}
	\caption{The coverage probability of the typical user. }
	\label{fig:fig4coveragewithris}
\end{figure}
Figs. \ref{fig:riselements} and \ref{fig:fig4coveragewithris} depict the coverage probability of the networks for various $N_r$. It is noteworthy that we reference Table \ref{Table:1}. Our observations indicate that the coverage probability of the typical user—specifically, the coverage probability of the proposed RIS-assisted network—increases as (i) the SNR threshold value $\tau$ decreases and (ii) the number of RIS elements $N_r$ increases. The coverage probability behavior concerning $N_r$ is explicitly delineated in Fig. \ref{fig:riselements}.

Upon analyzing Figs. \ref{fig:riselements} and \ref{fig:fig4coveragewithris}, we discern that for a higher SNR threshold, the addition of RIS elements has a marginal impact. In contrast, for a lower SNR threshold, the impact of adding more RIS elements becomes more pronounced. This underscores that if RIS-enabled applications necessitate a high SNR threshold, the inclusion of additional RIS elements does not significantly enhance the large-scale connectivity of the network. Conversely, in scenarios where RIS-enabled applications operate under low SNR conditions, it is crucial to recognize that the introduction of extra RIS elements significantly boosts the connectivity of the RIS-assisted network architecture.

\begin{figure}
	\centering
	\includegraphics[width=1\linewidth]{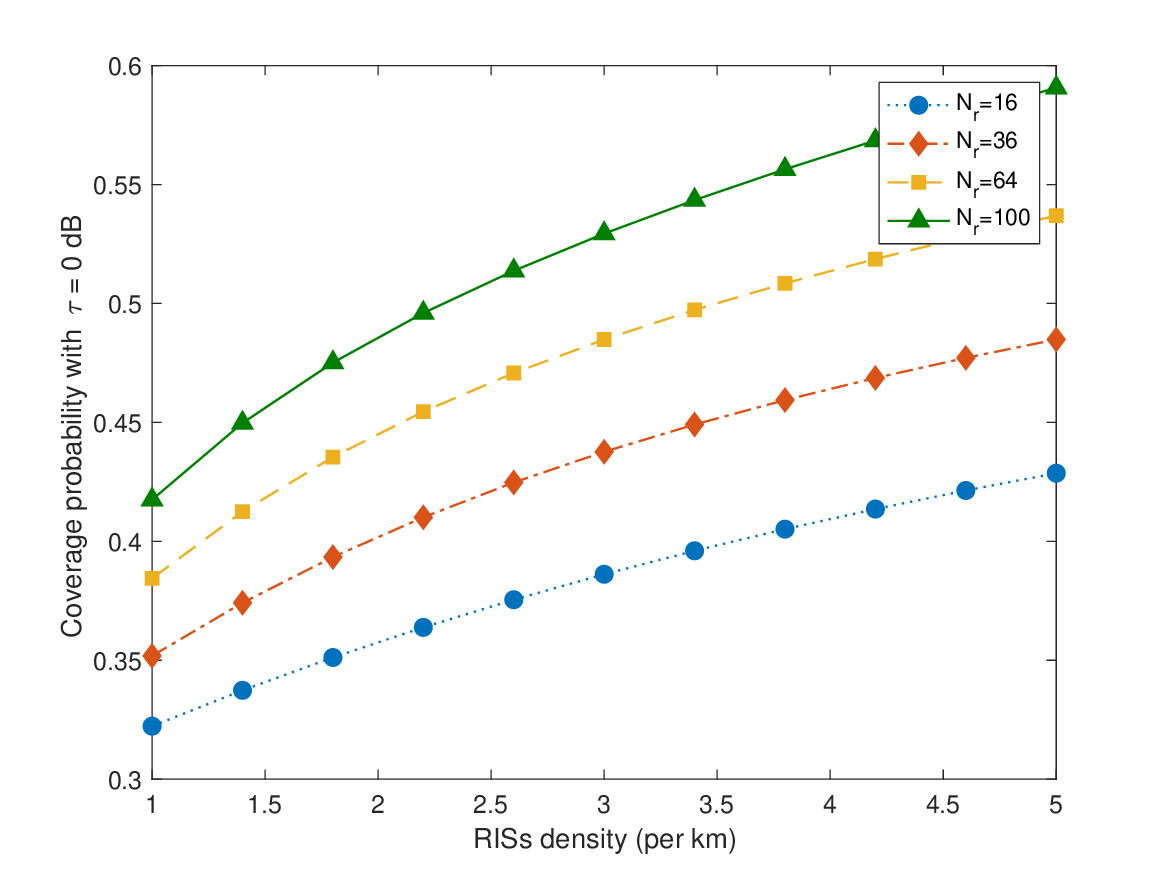}
	\caption{{The coverage probability of the network.  $\tau = 0$ dB.}}
	\label{fig:risdensityv2}
\end{figure}

\begin{figure}
	\centering
	\includegraphics[width=1\linewidth]{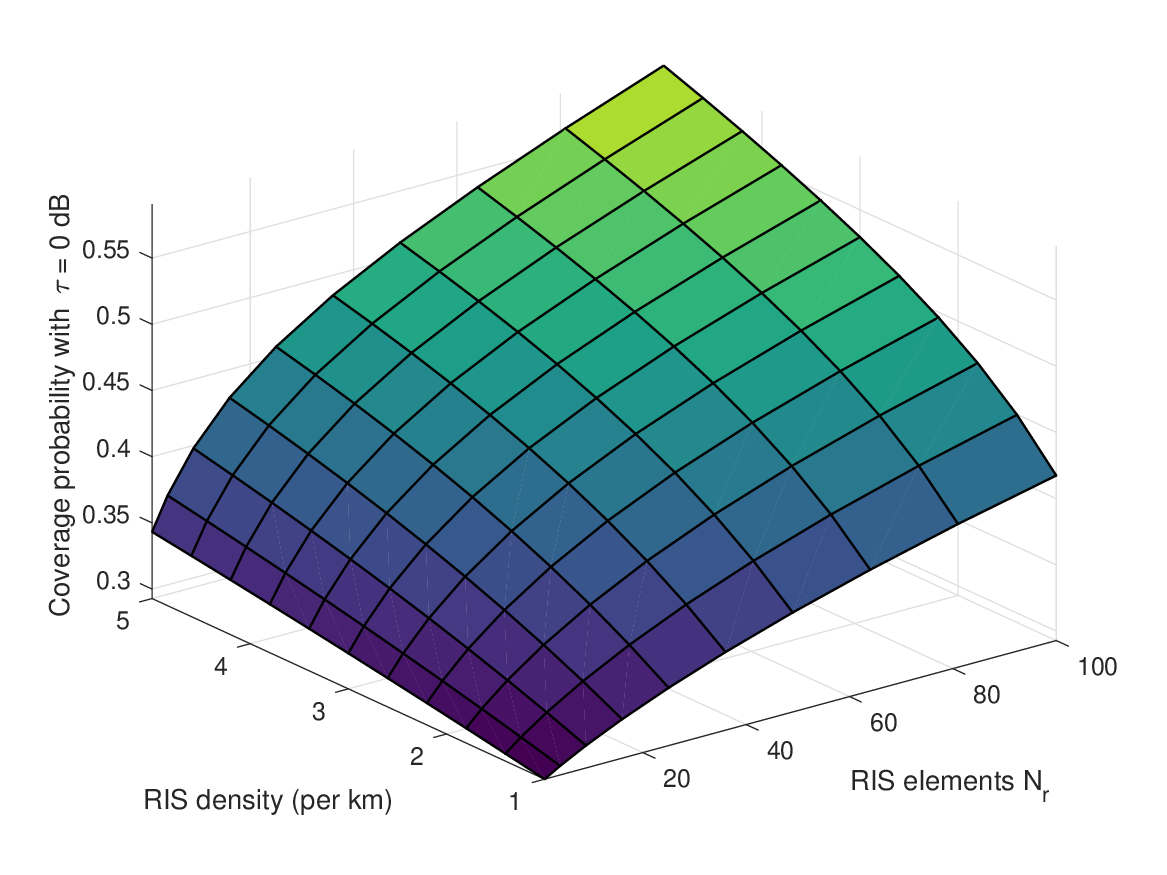}
	\caption{{The coverage probability of the network.  $\tau = 0$ dB.}}
	\label{fig:risdensityv2_2}
\end{figure}

\begin{figure}
	\centering
	\includegraphics[width=1\linewidth]{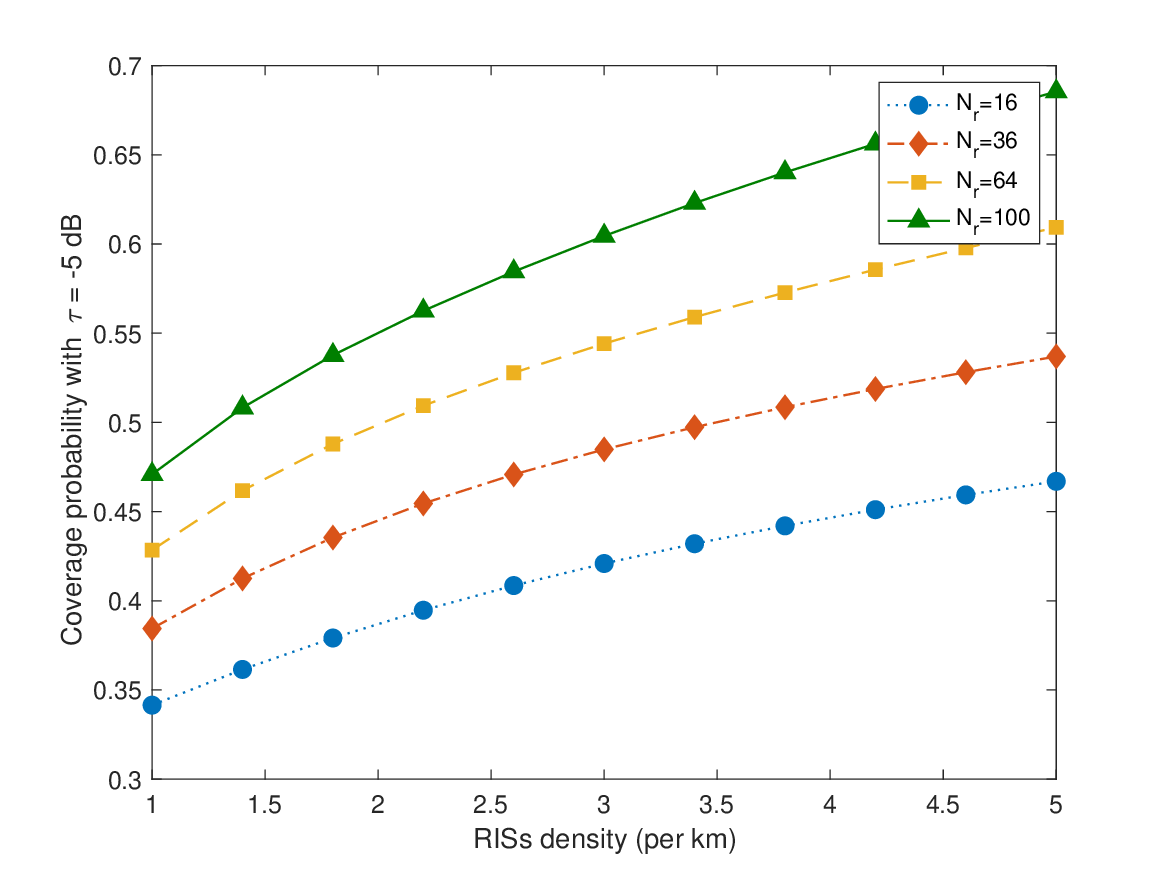}
	\caption{{The coverage probability of the network.  $\tau = -5$ dB.}}
	\label{fig:risdensityv2_5dB}
	
\end{figure}

\subsection{RIS Density}
Figs. \ref{fig:risdensityv2} and  \ref{fig:risdensityv2_2} illustrate the $0$ dB-threshold coverage probability as the function of RIS density. Note that the unit of RIS density $\mu$ is per km. Since RISs are strategically deployed around road areas and adding more RISs to the system will sharply improve the coverage of the network. As we deploy more RISs namely from $1$ to $5$ per km on average, the coverage probability increases. For $N_r=16$, the coverage probability grows from $0.32$ to $0.42$, which is about a $31$\% increment. For $N_r=100$, the coverage increases from $0.41$ to $0.59$, which is also about a $43$ \% increment.  

Note that the increase in the number of RIS elements has a more significant impact on the coverage probability when the RIS density is sufficiently high. Conversely, when the RIS density is not substantial, the influence of adding more RIS elements diminishes. This suggests that leveraging the network's crucial geometric features is vital for ensuring connectivity in RIS-assisted networks.

Fig. \ref{fig:risdensityv2_5dB} displays the coverage probability for SNR threshold $-5 $ dB which concerns a practical scenario that RIS-enabled applications leverage a relatively low SNR threshold to operate. The result shows that for low SNR threshold, deploying RISs to road areas gives the coverage probability between $0.32$ to $0.55$. Note that without any RIS, the coverage probability is $0.27$ and by deploying RISs to the system, we approximately have $20$\% ($1$ RIS per km and $N_r=16$) to $100 $ \% ($5$ RIS per km and $N_r=100$) coverage probability improvement. 



\begin{figure}
	\centering
	\includegraphics[width=1\linewidth]{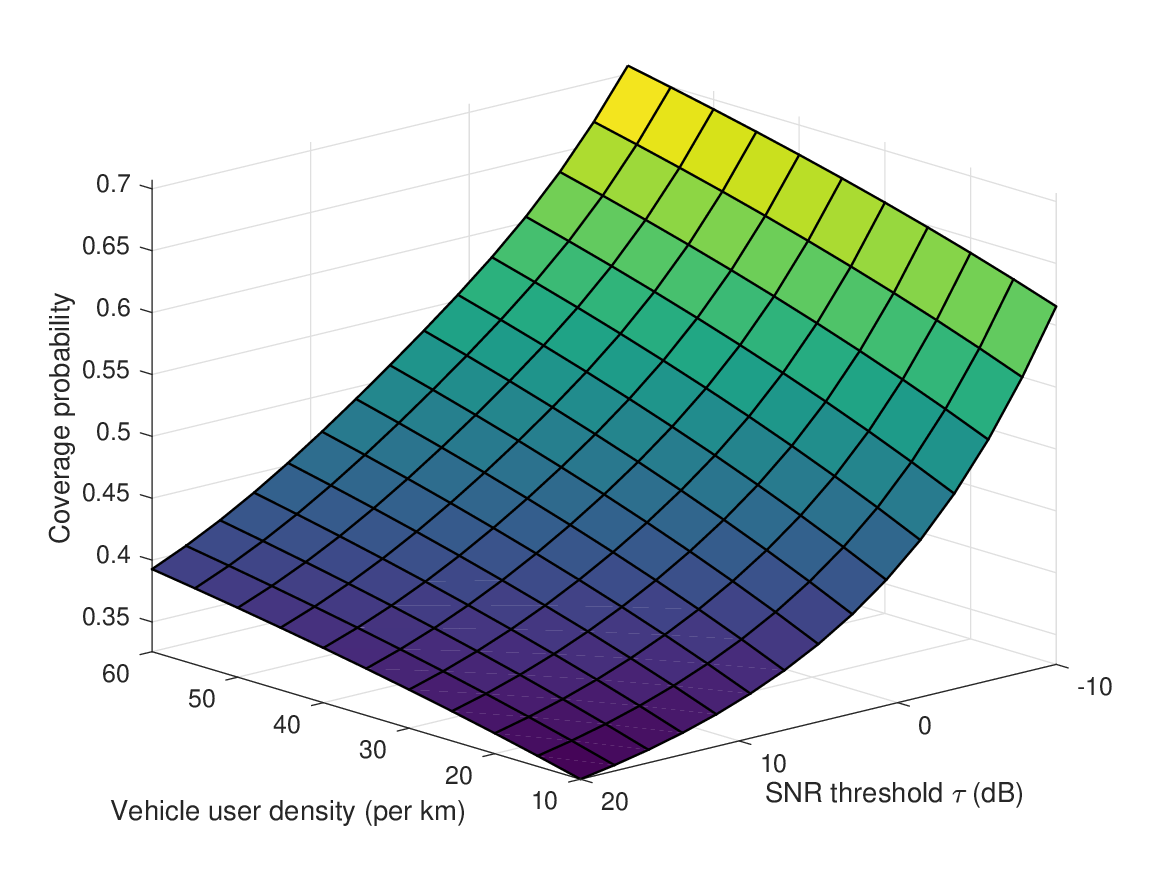}
	\caption{The coverage probability for user densities. $\lambda_2=300/\text{km}^2$}
	\label{fig:vehicledensity}
\end{figure}

\begin{figure}
	\centering
	\includegraphics[width=1\linewidth]{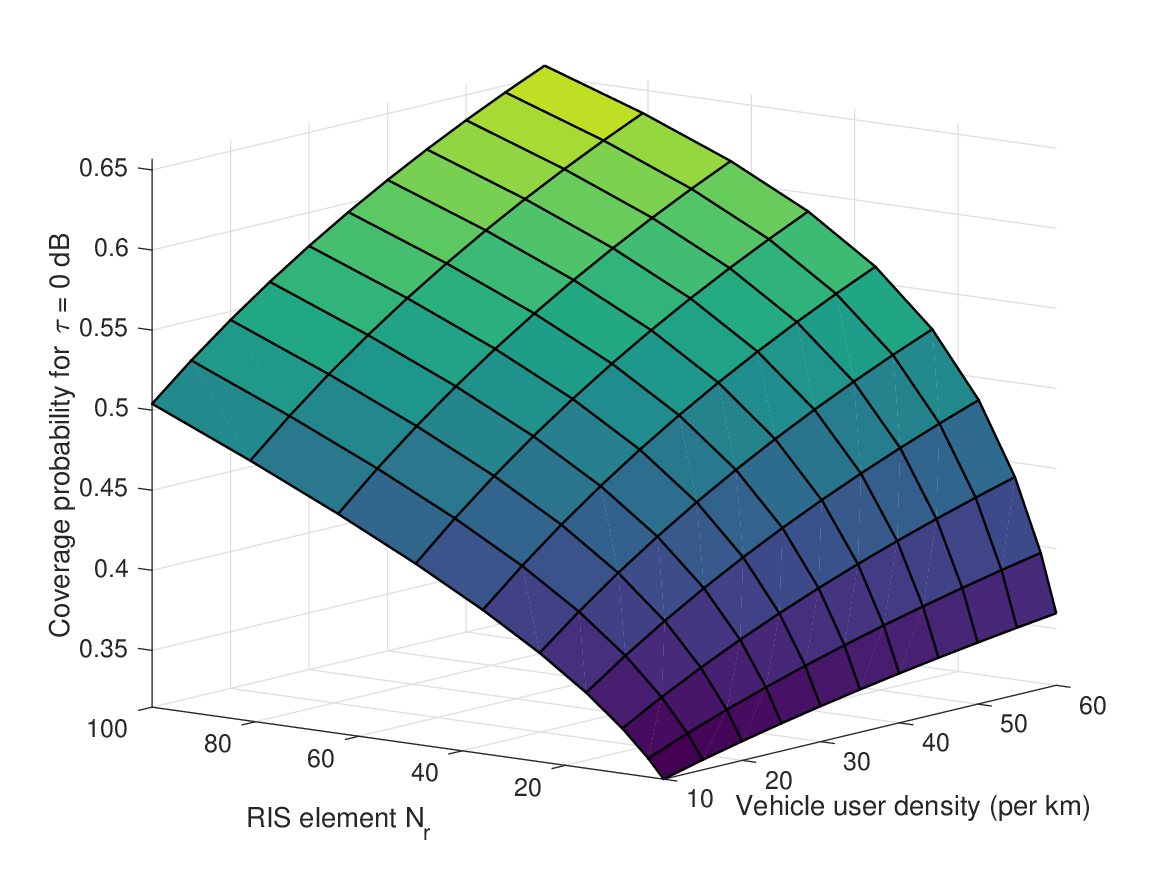}
	\caption{The coverage probability. $\lambda_2=200/\text{km}^2.$}
	\label{fig:vehicledensity2}
\end{figure}

%

\subsection{Vehicle Density and User Distribution}

Fig. \ref{fig:vehicledensity} shows the behavior of the coverage probability as the number of vehicle users grows when $\lambda_2=300/\text{km}^2$. The rest of parameters are in Table \ref{Table:2}. As the number of vehicle users increases, the coverage probability of the network monotonically improves. This is anticipated based on Theorem \ref{Table:2} since (i) the coverage probability of the typical user is given by the weighted summations of the coverage of the typical vehicle user and (ii) the coverage of the typical handset user and the coverage probability of the typical vehicle user receives a significant coverage benefit by RISs on road areas. The total vehicle density is given by the product of road density $\lambda$ and the vehicle user density $\nu$ and therefore, in the figure, the  vehicle spatial density is between $20/\text{km}^2$ and $120/\text{km}^2 $ with the experiment bench value at $50/\text{km}^2.$
 
Fig. \ref{fig:vehicledensity2} displays the coverage probability with the SNR threshold $\tau=0$ for various vehicle user densities and numbers of RIS elements. It demonstrates that the coverage probability of the network sharply improves as $N_r$ grows.  
\subsection{Geometric Difference: Path Loss}
\begin{figure}
	\centering
	\includegraphics[width=1\linewidth]{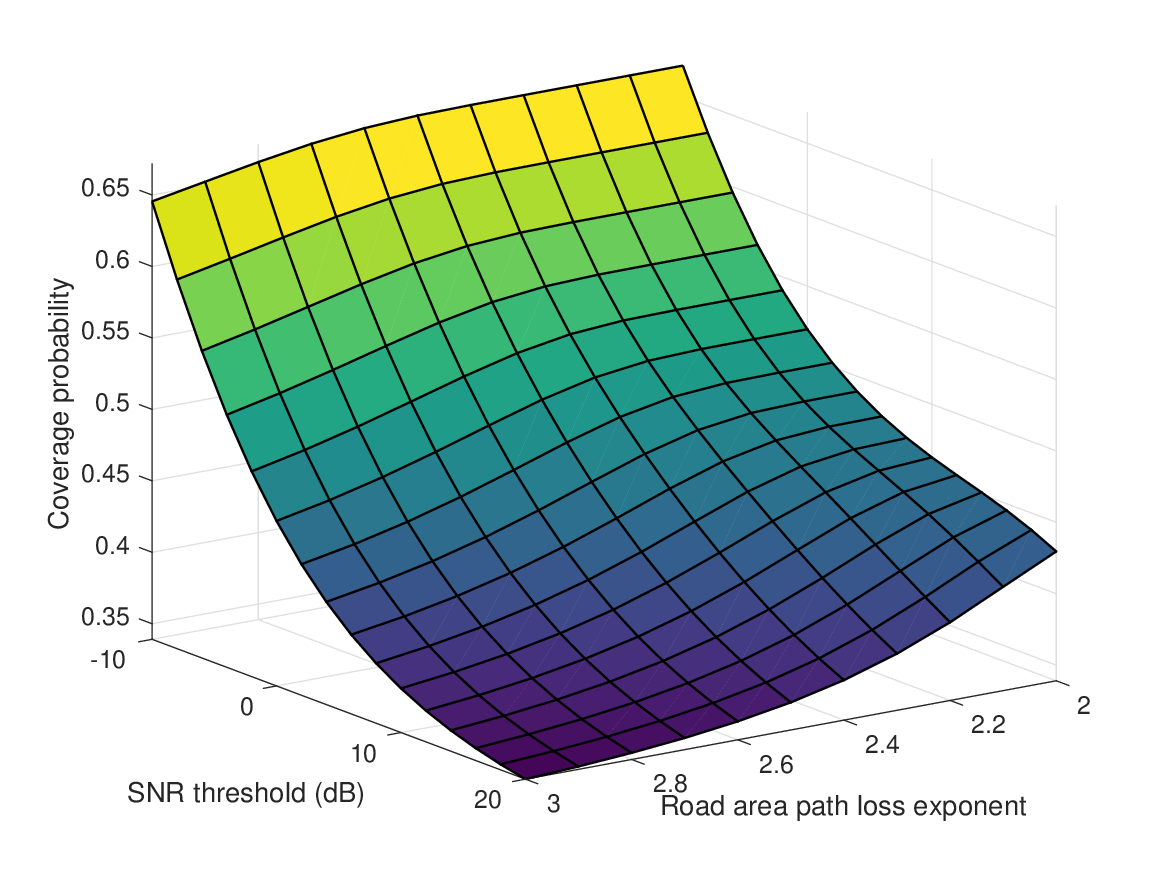}
	\caption{The coverage probability of the network for various  $\alpha_1$. $N_r=64.$}
	\label{fig:pathloss}
\end{figure}

Fig. \ref{fig:pathloss} illustrates the network's coverage probability for various values of $\alpha_{1}$. In this paper, we propose the deployment of RISs in road areas to fully leverage the favorable propagation and blockage conditions of these regions. Specifically, while the path loss exponent for urban areas is fixed at $\alpha_2=3.7,$ in road areas, it may vary between $2$ and $3$ with our benchmark value set at $\alpha_1=2.4$ based on \cite{38901}. As anticipated, a significant improvement in coverage probability is obtained by RISs as the path loss exponent of road areas, denoted as $\alpha_1$, transitions from $3$ to $2$. This shows that the potential coverage benefits offered by RISs become more evident as the path loss exponent difference between road and urban areas increases. Furthermore, the extent of coverage improvement is more pronounced for a lower SNR threshold $\tau$.

%

\subsection{Outage Gain from RISs}
Leveraging the coverage probability derived above, we express the benefit of employing RISs as an explicit function of the key network parameters. For simplicity, we use the outage probability defined as 1 minus the coverage probability. The motivation is based on the observation that the coverage probability derived in this paper is given by one minus a function, namely one minus the outage probability. By adopting the outage probability as a performance metric for evaluating the benefits of employing RISs, the outage gain will produce a surprisingly simple closed-form expression of network parameters. The outage gain is defined as the ratio of the outage without RISs to the outage with RISs as follows:
	\begin{equation}
	\mathcal{G} = \frac{\bP(\textnormal{outage without RIS})}{\bP(\textnormal{outage with RIS})}. 
\end{equation}
Using Theorem \ref{theorem:2}, the outage gain is given by 
	\begin{equation}
		\left(\frac{\lambda_l\nu}{\lambda_l\nu+\lambda_2} \cO_1+ 		\frac{\lambda_2}{\lambda_l\nu+\lambda_2}\cO_2 \right)^{-1},
	\end{equation}
	where  we have 
	\begin{align}
		\cO_1=&e^{-\mu\int_0^{c_1}\left\{1-f_{1}(t)\right\}\diff t}\cO_2,\nnb\\
		\cO_2=&\exp\left(-{\lambda_l}\!\!\int_0^{c_2}\!\!1-e^{-2\mu\int_0^{\sqrt{c_2^2-u^2}}1-f_{2}\left(\sqrt{u^2+t^2}\right)\diff t }\diff u\right).\nnb
	\end{align}
This outage gain directly illustrates the performance gain thanks to the RISs near road areas as the function of network parameters including $\lambda,\lambda_l,\mu,\nu,\lambda_2,\alpha_1,\alpha_2,\eta_{1}, \eta_2,$ and $\gamma.$ 
If the outage gain is one, it signifies that the network performances with and without RISs are nearly identical, indicating a negligible benefit from employing RISs. Conversely, an outage gain greater than one indicates that the network performance with RISs surpasses that without RISs, showcasing the coverage benefit conferred by reflective surfaces on roadsides.

\begin{figure}
	\centering
	\includegraphics[width=1\linewidth]{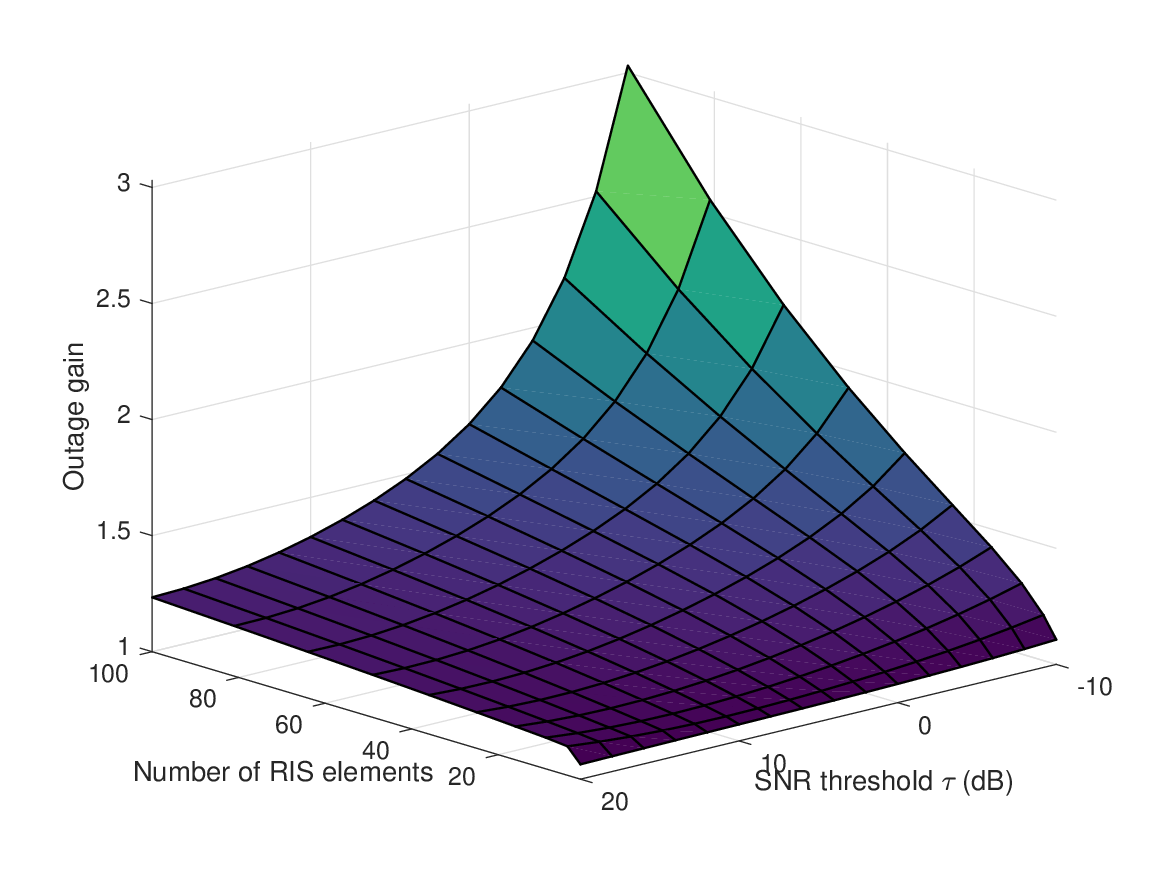}
	\caption{The gain of the RIS-assisted network w.r.t. $N_r$. }
	\label{fig:outagegain}
\end{figure}

Fig. \ref{fig:outagegain} shows the gain while varying the RIS density $\mu $ and the number of RIS elements $N_r$. The result shows that with RISs every $250 $ meters, when $N_r=1$ and $\tau=-20 $ dB, the outage gain is about $10\% $. In cases for higher $N_r$ or higher $\tau,$ we observe very high outage gains up to $200\%$ improvement. Note that the above obtained outage gain $\mathcal{G}$ is not directly translated into the coverage gain.  

\section{Conclusion}

RISs enhance network connectivity by providing reflected signals to users who may not be adequately served by base stations. Consequently, the locations of base stations, RISs, and network users are closely correlated. This work proposes an approach to exploit the distinctive local propagation and blockage associated with various links between these network elements.

We strategically deploy RISs along roads and analyze this deployment strategy by developing an analytical framework that captures the spatial correlation of network elements, the distinctive attenuation patterns in various areas, and the difference in blockage characteristics between urban and road areas. We define the coverage probability of the network and derive it as a closed-form expression that includes parameters such as RIS density, RIS elements, and user densities. Through numerical experiments and formulas, we demonstrate that the proposed approach successfully leverages differences in attenuation and blockage in the network, which could not be achieved simply by deploying RISs in random locations.

To focus on the geometric impact on network connectivity, we assume ideal RIS operation and no interference from RISs or base stations. Future work will include the analysis of these aspects, such as approximating total interference from other RISs or considering a certain energy threshold for operating RISs. These additional constraints may be evaluated under the presented framework, and it is evident that interference or a distinctive threshold may affect the connectivity of RIS-assisted cellular networks. However, the fundamental connectivity discussed in this paper is geometrically derived from the distributions of the network elements, making the coverage analysis still serve as the theoretical upper bound of RIS-assisted cellular networks.

\appendices
\section{Proof of Lemma \ref{Theorem:1}}\label{A:1}
Without any RISs, the coverage probability of the typical user is given by 
\begin{align*}
	&\bP_{\chi}^0(\text{cov} )=	1- \bP_{\chi}^0(\text{all signals w/ SNR} > \tau \text{ are blocked}). 
\end{align*}
Since signals come from base stations, we arrive at   
\begin{align}
	1-\bP_{\chi}^0(\text{cov} )
	&=\bE_{\chi}^0\left(\prod_{\Xi\in\Phi} \ind_{B_i}\ind_{\SNR_{i}>\tau }\right), \label{10}
\end{align}	
where we use the fact that points of Poisson base stations point process are independent.
In Eq. \eqref{10}, $\ind_{A}$ denotes an indicator function that takes one if the event $A$ occurs. In above, $\ind_{B_i}$ is one if the link from $X_i$ to the typical user is blocked and $\ind_{\SNR_{i}>\tau} $ is one if the average SNR of the link from $X_i$ to the typical user is greater than the threshold $\tau.$ 
The receive average signal power greater than $\tau$ is given by \[\frac{pG_t G_r(\lambda_c/4)^2}{L_pL_r N_oB_{W} F}{\|X\|}^{-\alpha_{2}}>\tau.\]
Since 	$\gamma= (p_tG_t G_r(\lambda_c/4)^2)/(L_pL_rN_0B_{W} F),$
we get 
\[\gamma \|X\|^{-\alpha_2}>\tau.\]

Now Eq. \eqref{10} is given by  
\begin{align}
	\bE_{\chi}^0\left(\prod_{X_i\in\Phi} \ind_{B_i}\ind_{\SNR_{i}>\tau }\right) & = 	\bE_{\chi}^0\left(\prod_{X_i\in\Phi} \ind_{B_i}\ind_{p/(\tau\|X_i\|^{\alpha_{2}})>\tau }\right)\nnb\\
	&=	\bE_{\chi}^0\left(\prod_{X_i\in\Phi} \ind_{B_i}\ind_{\|X_i\|<\left(\frac{\gamma}{\tau}\right)^{1/\alpha_{2}}}\right)\nnb \\
	&=	\bE_{\chi}^0\left(\prod_{X_i\in\Phi}^{\|X_i\|<\left(\frac{\gamma}{\tau}\right)^{1/\alpha_{2}}} \ind_{B_i}\right)\nnb\\
	&=\bE_{\chi}^0\left(\prod_{X_i\in\Phi}^{\|X_i\|<\left(\frac{\gamma}{\tau}\right)^{1/\alpha_{2}}} \bE[\ind_{B_i}|\Phi]\right),\nnb
\end{align}
where we use the assumption that users and base stations are independent, i.e., $\chi\independent \Phi $. 

By exploiting the probability generating functional of the Poisson point process of intensity $\lambda$ and then  employing the blockage probability expression, given by Eq. \eqref{eq:blockage}, we have 
\begin{align}
	\exp\left(-2\pi\lambda\int_{0}^{c_0}re^{-\frac{r}{\eta_{2}}}\diff r\right).
\end{align}

\section{Proof of Theorem \ref{theorem:2}}\label{A:2}
Recall the coverage probability of the typical user is 
\begin{align}
	\bP_{\chi}^0(\text{cov})
	=&\frac{\lambda_l\nu}{\lambda_l\mu+\lambda_2} \bP_{\chi_{1}}^0(\text{cov}) + \frac{\lambda_2}{\lambda_l\nu+\lambda_2} \bP_{\chi_{2}}^0(\text{cov})\label{eq:8} ,
\end{align}
where $\lambda_1 = \lambda_l\mu$  is the average number of vehicle user Cox point process in a unit area \cite{8419219} and $\lambda_2 $ is the average number of handset user Poisson point process in the same unit area.  Specifically, $\bP_{\chi_{1}}^0(\text{cov})$ is the coverage probability of the typical vehicle user and $\bP_{\chi_{2}}^0(\text{cov})$ is the coverage probability of the typical handset user.

The coverage probability of the typical vehicle user, denoted by $\bP_{\chi_1}^0(\textnormal{cov})$, is 
\begin{equation}
	\bP_{\chi_{1}}^0(\text{cov})= 1- \bP_{\chi_{1}}^0(\text{no LOS TX w/ SNR} > \tau).
\end{equation}
Since there will be direct signals from base stations and reflected signals from RISs, we have 
\begin{align}
	&\bP_{\chi_{1}}^0(\text{no LOS TX w/ SNR} > \tau)\nnb\\
	&=\bP_{\chi_{1}}^0(\text{no LOS BS w/ SNR} > \tau)\nnb\\
	&\hspace{6mm}\times \bP_{\chi_{1}}^0(\text{no LOS RIS w/ SNR} > \tau)\label{18}.
\end{align}
With regards to the first term of Eq. \eqref{18}, a base station at $X\in\bR^2$ provides the average SNR greater than $\tau $ if
\begin{equation}
	\gamma {\|X\|}^{-\alpha_{2}}>\tau. \label{11}
\end{equation}
The link from any base station $X$ to the typical user $O $ is the planar link. The above expression is equivalent to the following expression w.r.t. the distance from the origin to the base station.
\begin{equation}
	\|X-O\|<\left({\gamma/\tau}\right)^{1/\alpha_{2}}\equiv c_0.
\end{equation}
Using it, we have that \{the probability that there is no LOS base station with SNR $> \tau$\} is the same as \{the probability that all base stations within the distance $c_0$ are in NLOS\}. Therefore,  we get 
\begin{align}
	&\bP_{\chi_{1}}^0(\text{no LOS BS w/ SNR} > \tau)\nnb\\ &=\bE\left[\prod_{X_i\in\Phi}^{\|X_i\|<c_0}\ind_{B_i}\right] \nnb\\
	&=\!\exp\left(-2\pi\lambda\int_{0}^{c_0}re^{-\frac{r}{\eta_{2}}}\diff r\right),\label{13}
\end{align}
the same expression that we had for Lemma \ref{Theorem:1}. 

With regards to the second term of Eq. \eqref{18}, the typical user is a vehicle user and since the locations of vehicle users are given by a Cox point process exclusively on lines, there is a line that contains the typical vehicle user at the origin. In other words, under $ \bP_{\chi_{1}}^0 $---the Palm distribution of the vehicle user Cox point process---the typical vehicle user is at the origin and due to the conditional structure of the Cox point process,  there is a typical line $l_0 = l(r_0,\theta_0)$ \cite{8419219}. Since this typical line always contains the origin, we have $r_0\equiv 0 $. The typical line also makes an angle $\theta_0$ with the $x$-axis and we have $\theta_0\sim\text{Uniform}[0,\pi]$. 

For an RIS at $Y\in\bR^2$, the average reflected SNR is greater than $\tau$ if and only if 
\begin{align}
	{\gamma N_r\|X-Y\|^{-\alpha_2}{\|Y\|}^{-\alpha_1}}>\tau & \text{ for $Y \in l(0,\theta_0)$}, \label{20}\\
	{\gamma N_r \|X-Y\|^{-\alpha_2}{\|Y\|}^{-\alpha_2}}>\tau & \text{ for $Y \notin l(0,\theta_0)$}, \label{21}
\end{align}
Note that Eq. \eqref{20} is for the case where the RIS $Y$ is on the same line as the typical vehicle user and Eq. \eqref{21} is for the case where the RIS $Y$ is not on the same line as the typical vehicle user. For Eq. \eqref{20}, only the RISs at distances less than $c_1 $ may provide the average SNR greater than $\tau$ where $\gamma N_rc_1^{-\alpha_1}<\tau$ and
 \begin{equation}
	c_1 = (\gamma N_r^2/\tau)^{1/\alpha_1}.
\end{equation} 
Similarly, for Eq. \eqref{21}, only RISs  at distances less than $c_2$ can provide the average SNR greater than $\tau$ where $\gamma N_r c_2^{-\alpha_{2}}<\tau$ and 
\begin{equation}
	c_2=\left({\gamma N_r^2}/{\tau}\right)^{{1}/{\alpha_2}}.
\end{equation}

Conditionally on the location of the RIS $Y=y$, Eqs. \eqref{20} and \eqref{21} are given by 
\begin{align}
	&\|X-y\|<\left(\!\frac{\gamma N_r^2}{\tau{\|y\|}^{\alpha_1}}\!\right)^{\frac{1}{\alpha_2}}\!=\!\frac{{c_2}}{\|y\|^{\frac{\alpha_1}{\alpha_2}}}				 &\text{for $y \in l(0,\theta_0)$}, \label{22}\\
	&\|X-y\|<\left(\!\frac{\gamma N_r^2}{\tau{\|y\|}^{\alpha_2}}\!\right)^{\frac{1}{\alpha_2}}\!=\!\frac{{c_2}}{\|y\|} &\text{for $y \notin l(0,\theta_0)$}, \label{23}
\end{align}
respectively. Here, we define the exponent ratio as $	\rho = \alpha_1/\alpha_2.$

\begin{figure}
	\centering
	\includegraphics[width=.8\linewidth]{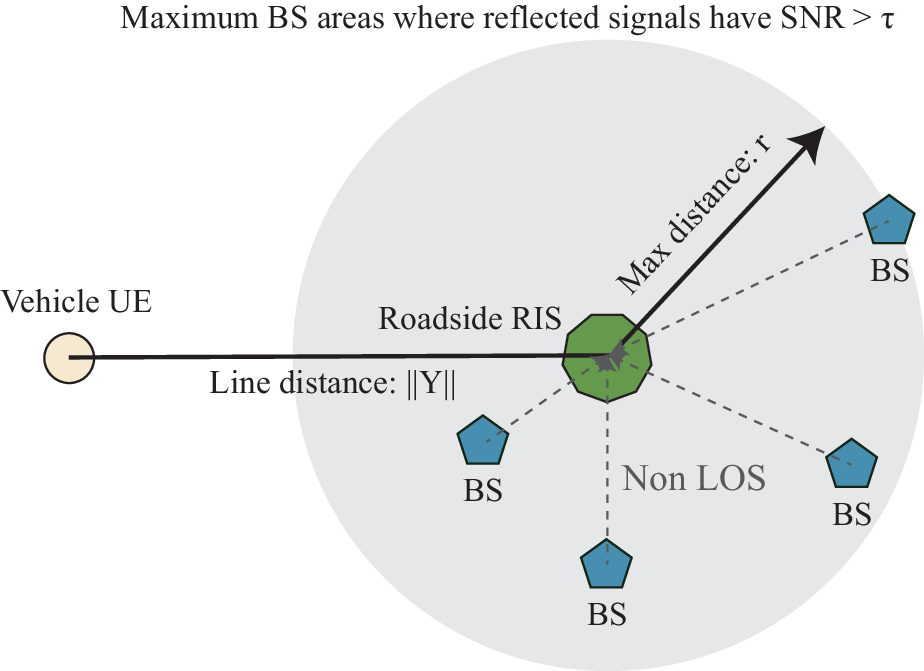}
	\caption{In Eqs. \eqref{20}, \eqref{22}, when the distance between typical user and RIS is $\|Y\|$, all the base stations at distance less than $c_2/\|Y\|^{\rho}$ from the RIS must be NLOS. In addition, the maximum value of $Y$ is given by threshold $\tau$, i.e., $\gamma\|y\|^{-\alpha_1}<\tau$.}
	\label{fig:nolosbswithrisonroad}
\end{figure}

Combining Eqs. \eqref{22} and \eqref{23}, we arrive at 
\begin{align}
	&\bP_{\chi_{1}}^0(\text{no LOS RIS w/ SNR} > \tau)\nnb\\
	&=\bP_{\chi_{1}}^0(!\exists X\text{ s.t. } \|X-Y\|<(\gamma N_r^2/\tau{\|Y\|}^{\alpha_{Y}})^{\frac{1}{\alpha_2}})\nnb\\
	&=\bE\left[\prod_{y_i\in\Psi+\phi_0}^{\|y_i\|<c_y}\ind_{B_i}\bE\left[\prod_{X_j\in\Phi}\ind_{B_{j|i}}\ind_{\|X_j-y\|<(\frac{\gamma N_r^2}{\tau\|y\|^{\alpha_{y}}})^{\frac{1}{\alpha_{2}}}}\right]\right]\nnb\\
	&=\bE\left[\prod_{y_i\in\phi_0}^{\|y_i\|<c_1}\ind_{B_i}\bE\left[\left.\prod_{X_j\in\Phi}\ind_{B_{j|i}}\ind_{\|X_j-y_{i}\|<c_2/\|y_{i}\|^{\rho}}\right|\phi_0\right]\right]\nnb\\
	&\hspace{3.7mm}\times \bE\left[\prod_{y_i\in\Psi^{0!}}^{\|y_i\|<c_2}\ind_{B_i}\bE\left[\left.\prod_{X_j\in\Phi}\ind_{B_{j|i}}\ind_{\|X_j-y_{i}\|<c_2/\|y_{i}\|}\right|\Psi\right]\right]\label{30}
\end{align}
where we use the Slivynak's theorem for the Poisson point process. In Eq. \eqref{30}, $\phi_0$ denotes the RIS point process on the typical line and $\Psi^{!0} $ is the total RIS Cox point process except the RISs on the typical line. 

%
%

Using the probability generating functionals of the point processes $\Phi$ and $\phi_0$, the first term of Eq. \eqref{30} is given by 
\begin{align}
	&\bE\left[\prod_{y_i\in\phi_0}^{\|y_i\|<c_1}\ind_{B_i}\bE\left[\left.\prod_{X_j\in\Phi}\ind_{B_{j|i}}\ind_{\|X_j-y_{i}\|<c_2/\|y_{i}\|^{\rho}}\right|\phi_0\right]\right]\nnb\\
	&=\bE\left[\prod_{y_i\in\phi_0}^{\|y_i\|<c_1}(1-e^{-\frac{t}{\eta_1}})\exp\left(-2\pi\lambda\int_{0}^{\frac{c_2}{t^{\rho}}} r e^{-\frac{r}{\eta_2}}\diff r\right)\right]\nnb\\
	&=\exp\left(-\mu\int_0^{c_1}1-f_{1}(t)\diff t \right),\label{35}
\end{align}
where we have defined a function $f_1(t)$ as follows: 
\begin{equation}
	f_{1}(t) = (1-e^{-\frac{t}{\eta_1}})\exp\left(-2\pi\lambda\int_{0}^{\frac{c_2}{t^{\rho}}} r e^{-\frac{r}{\eta_2}}\diff r\right).
\end{equation}

For the second term of Eq. \eqref{30}, we use the fact that $B_{j|i}$ is independent of everything else and the blockage probability over planar links is given by $ \bP(B_i) = 1-e^{-\|y\|/\eta_2}.$ We have 
\begin{align}
	&\bE\left[\prod_{y_i\in\Psi}^{\|y_i\|<{c_2}}\ind_{B_i}\bE\left[\prod_{X_j\in\Phi}^{\|X_j-y_{i}\|<{{c_2}}/\|y_i\|}\bE\left[\ind_{B_{j|i}}\right]\right]\right]\nnb\\
	&=\bE\left[\prod_{y_i\in\Psi}^{\|y_i\|<{c_2}}f_2(\|y\|)\right],
\end{align}
where 
\begin{align}
	f_{2}(\|y\|) = &(1-e^{-\|y\|/\eta_2})
	\exp\left(\left.-2\pi\lambda\int_{0}^{{{c_2}}/\|y\|}  r e^{-r/\eta_2}\diff r\right.\right).\nnb
\end{align}

By the definition of the Cox point process we write $\|y\| =\|T_{m;k}\| $ where $T_{m;k} $ is on the line $l_k$ and $T_{m;k}$ is the Poisson point process on the line. Furthermore, using the definition of the Poisson point process and its displacement theorem \cite{baccelli2010stochastic}, 
\begin{align}
	&\bE\left[\prod_{y_i\in\Psi}^{\|y_i\|<{c_2}}f_{2}(\|y\|)\right]\nnb\\
	& = \bE\left[\prod_{r_k\in\Xi}\bE\left[\prod_{T_{m;k}\in\phi_{k}}^{\|T_{m;k}\|<{{c_2}}} f_{2}\left(\|T_{m;k}\|\right)\right]\right]\nnb\\
	&=\bE\left[\prod_{r_k\in\Xi}\bE\left[\prod_{T_{n}\in\phi}^{r_k^2+T_n^2<{{c_2}}^2} f_{2}\left(\|r_k^2+T_n^2 \|\right)\right]\right]\nnb\\
	& = \bE\left[\prod_{r_k}\exp\left(-2\mu\int_0^{\sqrt{{c_2}^2-r_k^2}}1 - f_{2}\left(\sqrt{r_k^2+t^2}\right)\diff t \right)\right]\nnb\\
	&=\exp\left(-{\lambda_l}\!\int_0^{{c_2}}1-e^{-2\mu\int_0^{\sqrt{{c_2}^2-u^2}}\left(1-f_{2}\left(\sqrt{u^2+t^2}\right)\right)\diff t }\diff u \right).\label{17}
\end{align}
where we use the probability generating functional w.r.t. $\phi_k$ and then w.r.t. $\Xi.$ 
Therefore, combining Eq. \eqref{13}, \eqref{35}, and \eqref{17} gives the result in Eq. \eqref{18}.

On the other hand, the coverage probability of the typical handset user of Eq. \eqref{eq:8} is given by 
\begin{align}
	\bP_{\chi_{2}}^0(\text{cov}) = 1-\bP_{\chi_{2}}^0(\text{no LOS TX w/ SNR} > \tau).\label{14-1}
\end{align}
Under the Palm distribution, there exists a typical handset user at the origin. Note that in Eq. \eqref{14-1}, transmitters include all the base stations and RISs in the network. As a result, we have  
\begin{align}
	&\bP_{\chi_{2}}^0(\text{no LOS TX w/ SNR} > \tau)\nnb\\
	&=\bP_{\chi_{2}}^0(\text{no LOS BS w/ SNR} > \tau)\nnb\\
	&\hspace{3.7mm}\cdot \bP_{\chi_{2}}^0(\text{no LOS RIS w/ SNR} > \tau).\label{15-1}
\end{align}
The first term of Eq. \eqref{15-1} is given by 
\begin{equation}
	\exp\left(-2\pi\lambda\int_{0}^{c_0}re^{-\frac{r}{\eta_{2}}}\diff r\right).
\end{equation}
Regarding the second term of Eq. \eqref{15-1},  an RIS at $Y\in\bR^2$ provides the average SNR greater than $\tau$ if 
\begin{equation}
	{\gamma N_r^2\|X-Y\|^{-\alpha_2}{\|Y\|}^{-\alpha_2}}>\tau,\label{14}
\end{equation}
where $X$ is the location of the base station operating the RIS. 
For Eq. \eqref{14}, only RISs at distances less than $c_2$ may provide the average SNR greater than $\tau$ where $\gamma N_r^2 c_2^{-\alpha_2}>\tau$. For such RISs, Eq. \eqref{14} is given by  
\begin{align}
	\|X-y\|<\left(\frac{\gamma N_r^2}{\tau\|y\|^{\alpha_2}}\right)^{\frac{1}{\alpha_2}} = \frac{{c_2} }{\|y\|}.\label{42}
\end{align}
Finally, we have 
\begin{align}
	&\bP_{\chi_{2}}^0(\text{no LOS RIS w/ SNR} > \tau)\nnb\\
	&=\bP_{\chi_{2}}^0(!\exists X\text{s.t. }\|X_y-y\|<({{c_2}}/\|y\|)\nnb\\
	&=\bE_{\chi_{2}}^0\left[\prod_{y_i\in\Psi}^{\|y_i\|<{c_2}}\ind_{B_i}\bE\left[\left.\prod_{X_j\in\Phi}\ind_{B_{j|i}}\ind_{\|X_j-y\|<{{c_2}}/\|y\|}\right|\Psi\right]\right]\nnb\\
	&=\bE\left[\prod_{y_i\in\Psi}^{\|y_i\|<{c_2}}\ind_{B_i}\bE\left[\left.\prod_{X_j\in\Phi}\ind_{B_{j|i}}\ind_{\|X_j-y\|<{{c_2}}/\|y\|}\right|\Psi\right]\right]\nnb\\
	&=\exp\left(-{\lambda_l}\!\int_0^{{c_2}}1-e^{-2\mu\int_0^{\sqrt{{c_2}^2-u^2}}\left(1-f_{2}\left(\sqrt{u^2+t^2}\right)\right)\diff t }\diff u \right).\label{43}
\end{align}
Finally,  multiplying Eqs.  \eqref{42} and \eqref{43} gives the result in Eq. \eqref{15-1}. 

\section*{Acknowledgment}
This work was supported partly by Institute of Information \& communications Technology Planning \& Evaluation (IITP) grant funded by the Korea government(MSIT)  (RS-2023-00216221, Development of service coverage extension technologies for 5G-Advanced mobile communications based on reconfigurable intelligent surface) and by the National Research Foundation of Korea(NRF) grant funded by the Korea government(MSTI) (No RS-2024-00334240).

\bibliographystyle{IEEEtran}
\bibliography{ref}

\end{document}